\documentclass[11pt,a4paper]{article}

\usepackage{jheppub}
\usepackage[T1]{fontenc}
\usepackage{lmodern}
\usepackage{amsmath,amssymb,mathtools,mathrsfs,empheq}
\usepackage{caption}
\usepackage[dvipsnames]{xcolor}
\usepackage{stmaryrd}
\usepackage{mdframed}

\newcommand\nn{\nonumber}

\renewcommand{\v}{\mathbf{v}}
\renewcommand{\u}{\mathbf{u}}
\newcommand{\w}{\mathbf{w}}
\newcommand{\N}{\mathcal{N}}
\newcommand{\Z}{\mathbb{Z}}
\newcommand{\I}{\mathcal{I}}
\newcommand{\Q}{\mathcal{Q}}
\newcommand{\beps}{\boldsymbol{\varepsilon}}
\newcommand{\A}{\mathcal{A}}
\newcommand{\B}{\mathcal{B}}
\newcommand{\R}{\mathbb{R}}
\newcommand{\hatz}{\hat{\mathbf{z}}}
\newcommand{\C}{\mathcal{C}}
\newcommand{\V}{\mathcal{V}}
\renewcommand{\b}{\mathbf{b}}
\renewcommand{\P}{\mathcal{P}}
\renewcommand{\H}{\mathcal{H}}

\newcommand{\modker}{\Upsilon}

\global\mdfdefinestyle{exampledefault}{%
leftmargin=0.1cm,rightmargin=0.1cm
}

\newcommand{\para}{\paragraph{}}

\title{Giant Gravitons and Volume Minimisation}

\author[a,b]{Heng-Yu Chen,}
\author[c]{Nick Dorey,}
\author[d]{Sanefumi Moriyama,}
\author[e]{Rishi Mouland,}
\author[f]{and Canberk \c{S}anl\i}

\affiliation[a]{Department of Physics, National Taiwan University, Taipei 10617, Taiwan}
\affiliation[b]{Center for Gravitational Physics and Quantum Information, Yukawa Institute for Theoretical Physics, Kyoto University, Kitashirakawa Oiwakecho, Sakyo-ku, Kyoto 606-8502, Japan }
\affiliation[c]{DAMTP, Centre for Mathematical Sciences, University of Cambridge, Wilberforce Road, Cambridge CB3 0WA, U.K.}
\affiliation[d]{Department of Physics, Graduate School of Science, Osaka Metropolitan University, Sumiyoshi-ku, Osaka, Japan 558-8585}
\affiliation[e]{Blackett Laboratory, Imperial College London, Prince Consort Road, London, SW7 2AZ, U.K. }
\affiliation[f]{CEICO, Institute of Physics of the Czech Academy of Sciences, Na Slovance 2, 182 00 Prague 8, Czech Republic}

\emailAdd{heng.yu.chen@phys.ntu.edu.tw}
\emailAdd{n.dorey@damtp.cam.ac.uk}
\emailAdd{moriyama@omu.ac.jp}
\emailAdd{r.mouland@imperial.ac.uk}
\emailAdd{sanli@fzu.cz}

\abstract{
We establish a precise correspondence between the giant graviton expansion of the superconformal index of field theories in $D\leq 4$, and the master volume formalism of Gauntlett, Martelli and Sparks (GMS) which determines the near horizon geometries of certain BPS black holes and black strings in supergravity. We focus on 4d $\N=1$ superconformal field theories arising on the world volume of $N$ D3 branes placed at the tip of a cone over a toric Sasaki-Einstein manifold SE$_{5}$, the simplest example of which is $S^5$, corresponding to $\N=4$ super-Yang-Mills. 
The giant graviton expansion realises the superconformal index as the sum of contributions from wrapped D3 branes in the dual AdS$_{5}\times \text{SE}_{5}$. We argue that, for large wrapping numbers, the asymptotics of each such contribution is governed by the {\em master volume} of a particular metric deformation of $\text{SE}_5$ (suitably fibred over $S^{3}$). 
In particular, the wrapping numbers of a generic giant graviton configuration are identified with K\"{a}hler moduli of the corresponding metric. 
We further show that at large $N$ the entropy function of the relevant AdS$_5\times \text{SE}_5$ BPS rotating black hole is recovered by extremising over these moduli. Our results suggest that the complex Euclidean geometries corresponding to rotating BPS black holes in AdS$_{5}$ are determined by a close analogue of GMS volume minimisation, and that conversely, the off-shell geometries considered in such minimisation procedures should be understood as the near-horizon geometries of back-reacted giant gravitons.
We present analogous results for 3d $\N=2$ theories holographically dual to M-theory on AdS$_4\times \text{SE}_7$.
} 

\begin{document}
	
\maketitle 

\section{Introduction}
\label{sec: intro}

Supersymmetric black holes in anti-de-Sitter space offer a uniquely tractable setting for the study of quantum gravity. In particular, the AdS/CFT correspondence provides a microscopic interpretation of Bekenstein-Hawking entropy in terms of the degeneracy of BPS states in the CFT. Indeed the gravitational formula for the entropy can be reproduced by counting these states using the superconformal index of the dual field theory. Despite much subsequent progress, a 
more detailed understanding of the map between the gravitational and CFT dynamics underlying this agreement remains elusive. In this paper we will uncover a new aspect of this map, by demonstrating a precise relation between two previously unconnected strands of recent progress on either side of the AdS/CFT correspondence. 

\para 
On the gravity side, the near-horizon geometry of BPS black holes in supergravity can sometimes be determined by stationarising an action over a suitable variational family of supersymmetry-preserving metrics\footnote{Note that this connection has only been established for static, magnetically charged black holes \cite{Gauntlett:2019roi} and black strings \cite{Gauntlett:2018dpc} whose entropy is captured by a topologically twisted index in the dual field theory. We emphasise that, so far, no such procedure is known for determining near horizon geometry of the rotating AdS black holes corresponding to the field theory superconformal index.}. The action is directly related to the volume of a corresponding internal space which depends on some set of variational parameters\footnote{The action also includes additional source terms corresponding to various fluxes through cycles of the geometry.}. In the following, we will use the recently discovered \textit{giant graviton} expansion \cite{Imamura:2021ytr,Gaiotto:2021xce} of the superconformal index to show that a closely related variational problem appears in the boundary field theory. Our results suggest that the near-horizon geometries of rotating $\frac{1}{16}$-BPS black holes in AdS are similarly determined by variation over a family of supersymmetry-preserving metrics, and furthermore that each member of this family corresponds to a back-reacted supergravity configuration sourced by a specific configuration of giant gravitons.                
\para
We will focus on the large family of $\mathcal{N}=1$ superconformal field theories in four dimensions which arise as worldvolume theories of $N$ D3 branes placed at the tip of a cone over a toric Sasaki-Einstein manifold $\text{SE}_5$ and on the dual IIB string backgrounds of the form AdS$_{5}\times \text{SE}_5$. Both the field theory Lagrangian and the dual geometry are determined in terms of the {\em toric data} of the cone. This is specified by a set of vectors $\v_I\in \Z^3$ with $ I=1,\dots,D$, for some $D\ge 3$. We also present analogous results for 3d $\mathcal{N}=2$ theories holographically dual to M-theory on AdS$_{4}\times \text{SE}_{7}$, but will focus on the four-dimesional case for the remainder of this introductory section. 
\para
On the field theory side of the correspondence, our main focus is on the superconformal index $\mathcal{I}(\omega_{i}, \Delta_{I})$ which contains information about the spectrum of BPS states graded by conserved charges. The index depends on chemical potentials $\omega_{i}$, $i=1,2$, for angular momenta and $\Delta_{I}$ for global symmetries\footnote{The field theories in question have $n_{M}=3$ mesonic $U(1)$ symmetries corresponding to the toric isometries of $\text{SE}_5$. They also have $n_{B}={\rm dim}H_{3}(\text{SE}_5)$ baryonic $U(1)$ symmetries corresponding to non-contractible three-cycles in spacetime. Here we introduce an index $I=1,2,\ldots,D=n_{M}+n_{B}$ which runs over both types of global symmetry. It coincides with the index labelling the toric vectors $\v_I\in \Z^3$ introduced above. }.
The giant graviton expansion provides a new representation of the superconformal index as an infinite series of the form, 
\begin{align}
  \I(\omega_i,\Delta_I) = \I_\infty(\omega_i,\Delta_I)\sum_{n_1,\dots,n_D=0}^\infty e^{-N n_I \Delta_I} \I_{(n_1,\dots,n_D)}(\omega_i,\Delta_I)\ .
\label{eq: GG expn intro}
\end{align}
This expression has a natural interpretation in the dual string theory background. The prefactor  $\I_{\infty}$ 
coincides with the index of classical supergravity on AdS$_{5}\times \text{SE}_5$, and dominates the index at large $N$, for a suitable scaling of chemical potentials with $N$ such that charges are $\mathcal{O}(N^0)$. The remaining terms account for the contribution of {\em giant gravitons}, or D3 branes wrapped on supersymmetry preserving three-cycles in $\text{SE}_5$. These are labelled by a set of non-negative integers $(n_{1},\ldots,n_{D})$. For each value of these wrapping numbers, the index includes a factor $\I_{(n_1,\dots,n_D)}$ that accounts for the BPS states of open strings ending on the wrapped D-branes. We will refer to this factor as the {\em giant graviton index}. Imamura and collaborators have proposed an explicit formula \cite{Fujiwara:2023bdc} for the giant graviton index as a suitable analytic continuation of the index of a certain 4d--2d coupled CFT corresponding to the world-volume gauge theory of the intersecting wrapped branes. Importantly, because of the analytic continuation, giant graviton configurations contribute with alternating signs beyond those corresponding to the factor of $(-1)^{F}$ in the definition of the index. This indicates a vast redundancy in the stringy description of these states which we comment on below. 

\para
Our interest lies in the behaviour of the giant graviton index $\I_{(n_1,\dots,n_D)}$ at large wrapping numbers $n_I \gg 1$. We are motivated to do so, since at large $N$ (and fixed chemical potentials) the index $\I(\omega_i,\Delta_I)$ is dominated by terms in the giant graviton expansion with $n_I \sim N\gg 1$. We will show that the large $n_I$ asymptotics of $\I_{(n_1,\dots,n_D)}$ are determined by the \textit{master volume} of a certain auxiliary geometry, which we now describe.

\para
The topological space $Y_{5}\simeq \text{SE}_5$ admits a family of Sasakian metrics parametrised by a choice of Reeb vector field. In the seminal work of Martelli, Sparks and Yau \cite{Martelli:2006yb}, the Sasaki-Einstein metric on $\text{SE}_5$ is determined by minimising the volume of $Y_{5}$ with respect to the choice of Reeb vector field. This variational family was further enlarged by Gauntlett, Martelli and Sparks (GMS) \cite{Gauntlett:2018dpc} to include additional transverse K\"{a}hler parameters\footnote{$Y_{5}$ is realised as a Reeb fibration over a conformally K\"{a}hler manifold. The new parameters parametrise the K\"{a}hler class of the base. The resulting metrics are no longer Sasakian and further fibration of the corresponding $Y_{5}$ over $\mathbb{CP}^{1}$ give rise to so-called GK geometries. The latter form a variational family for the near-horizon geometry of black strings in AdS$_{5}\times \text{SE}_5$.}. 
The resulting volume of $Y_{5}$ as a function of these parameters is known as the {\em master volume} function. For toric $Y_{5}$, the choice of Reeb vector field corresponds to the choice of a vector $\b\in \R^3$ and the transverse K\"{a}hler class is specified by $D$ additional real parameters $\lambda_{I}$. The master volume function $\V(\b,\lambda_I)$ is then uniquely determined in terms of the toric data $\{\v_I\in \mathbb{Z}^{3}: I=1,\ldots,D\}$ as given in (\ref{eq: AdS5 master volume}) below. Note that the parameterisation $(\b,\{\lambda_I\})$ of five-dimensional geometries exhibits a two-parameter redundancy, such that for each $\{\lambda_I\}$ there exists a two-parameter family of alternative $\{\lambda_I'\}$ such that $\V(\b,\lambda_I) = \V(\b,\lambda_I')$. 

\para
To summarise, the internal space SE$_5$ of the asymptotic bulk geometry can be realised as a special point in a variational family of metrics on the same topological space, parameterised by the Reeb vector $\b$ and K\"ahler parameters $\lambda_I$. To each member of this family we can associate a master volume $\V(\b,\lambda_I)$.
 
\para 
The key proposal of this paper is that the asymptotics of the giant graviton index at large winding number are governed by the master volume of another metric in this same variational family. Explicitly, 
for $n_{I}\gg 1$, we have    
\begin{align}
  \boxed{\phantom{\bigg(}\log \I_{(n_1,\dots,n_D)}(\omega_i,\Delta_I) \sim -\frac{\omega_1 \omega_2}{(2\pi)^3} \,\V(\b,\lambda_I)\phantom{\bigg)}}\qquad \text{with}\qquad  \b = \sum_I \Delta_I \v_I,\quad \lambda_I = n_I\ .
\label{eq: main result intro}
\end{align} 
In Section \ref{sec: AdS5}, we show that (\ref{eq: main result intro}) holds for generic toric theories provided a certain residual matrix integral scales as order $(n_I)^0$ in the large $n_I$ limit\footnote{More precisely the logarithm of this integral must grow slower than $\mathcal{O}(n_I^2)$.}. We are able to verify this property rigorously only in the case of $S^{5}$. Interestingly, the gauge symmetry of the master volume mentioned above is mapped by our proposed relation to a similar redundancy in the index, related to the `simple sum' formulation of the giant graviton index \cite{Fujiwara:2023bdc}. In the following, we present evidence that for a generic toric theory, the asymptotic (\ref{eq: main result intro}) holds (for some values of chemical potentials) at at least one point on each gauge orbit. 

\para  
The identification of brane wrapping numbers $n_{I}$ and the K\"{a}hler moduli $\lambda_{I}$ is very suggestive of the geometric transition \cite{Cachazo:2001jy} in topological string theory and suggests that the corresponding metric on $Y_{5}$ (further fibered over $S^{3}$) should be related to the back-reacted geometry sourced by the giant gravitons. We discuss this interpretation further in the final section of the paper. We also note that the relation (\ref{eq: main result intro}) has important precursors in the literature. 
 It is strongly reminiscent of the relation between the asymptotics of the Hilbert series of a Calabi-Yau cone and the Sasakian volume of the link uncovered in \cite{Martelli:2006yb} (see also \cite{Martelli:2006vh} for further developments). It would be interesting to find the precise link between these two relations, perhaps by considering a limit or specialisation of our results for which the superconformal index of the 4d theory reduces to the Hilbert series of its Higgs branch. 
\para
Finally we turn our attention to the large $N$ regime. We would like to verify that at leading order in this regime, $\log \I(\omega_i, \Delta_I)$ coincides with (minus) the Euclidean on-shell action of a suitable asymptotically AdS$_{5}\times \text{SE}_{5}$ supersymmetric black hole \cite{Cabo-Bizet:2018ehj}. This is otherwise known as the \textit{entropy function}, since the microcanonical black hole entropy can be recovered by a suitable Legendre transform. For $N\gg 1$, we show that the series (\ref{eq: GG expn intro}) is dominated by high order terms where the sum over wrapping numbers $\{n_{I}\}$ can be replaced by an integral over continuous variables $\{\lambda_{I}\}$ and the giant graviton index $\I_{(n_1,\dots,n_D)}$ can be replaced by its asymptotic form (\ref{eq: main result intro}). The resulting integral can be evaluated by a saddle-point approximation, with saddle scaling as $\lambda_I\sim N$, to arrive at   
\begin{align}
  \log \I(\omega_i,\Delta_I) \sim  \frac{ N^2}{12\omega_1\omega_2}\sum_{I,J,K}C_{I,J,K}\Delta_I \Delta_J \Delta_K\ , \qquad C_{I,J,K}:= |\v_I \cdot (\v_J \times \v_K)|\ .
  \label{eq: total index asymptotic intro}
\end{align}
This result matches precisely the expected entropy function of the corresponding BPS black hole, as computed in field theory \cite{Choi:2023tiq,Amariti:2019mgp,Lanir:2019abx,Benini:2020gjh,Cabo-Bizet:2018ehj,Cabo-Bizet:2019osg,Cabo-Bizet:2020nkr}, while also matching all known relevant supergravity results.

\para
Recasting the computation of black hole entropy in this way, the saddle point analysis of the resulting integral requires us to stationarise the master volume function with respect to the K\"{a}hler parameters $\{\lambda_{I}\}$ in the presence of source terms. The Legendre transformation to recover the entropy similarly corresponds to finding the stationary point with respect to variations of the toric Reeb vector $\mathbf{b}$, along with the angular momentum chemical potentials $\omega_1,\omega_2$ and the $(D-3)$ other independent linear combinations of the chemical potentials. As we discuss in Section 4, this strongly suggests the existence of a geometrical variational principle for the near-horizon geometry of rotating AdS black holes, generalising the one discovered by GMS for static, magnetically charged black holes and black strings. A significant difference\footnote{Another, more superficial, difference is that, in the analysis of \cite{Gauntlett:2019roi,Gauntlett:2018dpc}, the K\"{a}hler parameters are eliminated via a set of constraints related to flux quantization rather than variational equations of motions. Note however, as in our case, these constraints can easily be reformulated as the stationary condition for an associated action. We discuss this further in Section \ref{sec: discussion}.} to the static case is that rotating black holes correspond to complex stationary points of the volume corresponding to complex Euclidean geometries. This is of course expected, since a Wick-rotated rotating black hole is generically described by a complex metric. Our results also suggest a physical interpretation for the full variational family of metrics as back-reacted supergravity solutions sourced by the full set of giant graviton configurations. It would be interesting to establish a similar interpretation in the static, magnetically-charged case.     
\para
This paper is organised as follows. In Section \ref{sec: AdS5}, we study the AdS$_5$/CFT$_4$ case. We first introduce the basic elements of toric geometry, and state the master volume of the 5-manifold $Y_5$. We then change gears and study the superconformal index of the $\N=1$ SCFT dual to AdS$_5\times \text{SE}_5$. We outline the derivation of the large $n_I$ asymptotics of the giant graviton index $\I_{(n_1,\dots,n_D)}$, with details relegated to Appendix \ref{app: asymptotics}, and thus establish the main proposal (\ref{eq: main result intro}). We then consider the large $N$ limit, and by saddle-point approximation of the sum over $n_I$, recover the large $N$ asymptotic (\ref{eq: total index asymptotic intro}). Some minor details of this calculation are relegated to Appendix \ref{app: gauge orbit}. We finally discuss the extent to which we might think of a given BPS black hole as a collection of giant gravitons. Next, in Section \ref{sec: AdS4} we do it all over again, now in the context of 3d $\N=2$ theories dual to AdS$_4\times \text{SE}_7$. We present the relevant master volume formula, and use it to conjecture the large $n_I$ asymptotic of the corresponding giant graviton index. We explicitly verify this conjecture in the case SE$_7 = S^7$. Finally, in Section \ref{sec: discussion} we provide some discussion on how to interpret the quantitative result (\ref{eq: main result intro}) and its analogue in AdS$_4$, and what they mean for the relationship between giant gravitons and the class of variational geometries.

\section{The AdS$_5\times \text{SE}_5$ case}
\label{sec: AdS5}

We first study 4d $\N=1$ superconformal gauge theories that are dual to IIB string theory on an asymptotically AdS$_5\times \text{SE}_5$ spacetime. Such a scenario arises in the decoupling limit of $N$ D3-branes placed at the tip of a cone over SE$_5$. 

%

\subsection{Toric geometry and the master volume}\label{subsec: AdS5 master volume}

Our first task is to briefly review the construction of a Sasaki-Einstein manifold SE$_5$. In fact, our interest will be in a broader family of geometries, of which some SE$_5$ is a special member. We will be particularly interested in the associated volume of these geometries. This subsection constitutes a review of the basic construction of \cite{Gauntlett:2018dpc}\footnote{See also \cite{Martelli:2005tp}, as well as \cite{Martelli:2023oqk} for a contextualisation within equivariant localisation.}, which contains many more details than we include here.

\para
We start with a toric K\"ahler cone in 3 complex dimensions, which is additionally \textit{Gorenstein}, meaning it admits a global holomorphic $(3,0)$-form. This is a K\"ahler metric $X_6$ in six real dimensions, with a conical form
\begin{align}
  ds_6^2 = dr^2 + r^2 ds_5^2\ ,
\end{align}
that is invariant under a $U(1)^3$ isometry. The cone $X_6$ comes with a distinguished vector field known as the \textit{Reeb vector}, whose components in the basis of $U(1)^3$ vector fields  we denote $\b=(b^1,b^2,b^3)\in \R^3$.
\begin{center}\vspace{-1em}
\begin{minipage}{0.84\textwidth}
\centering
\includegraphics[width=80mm]{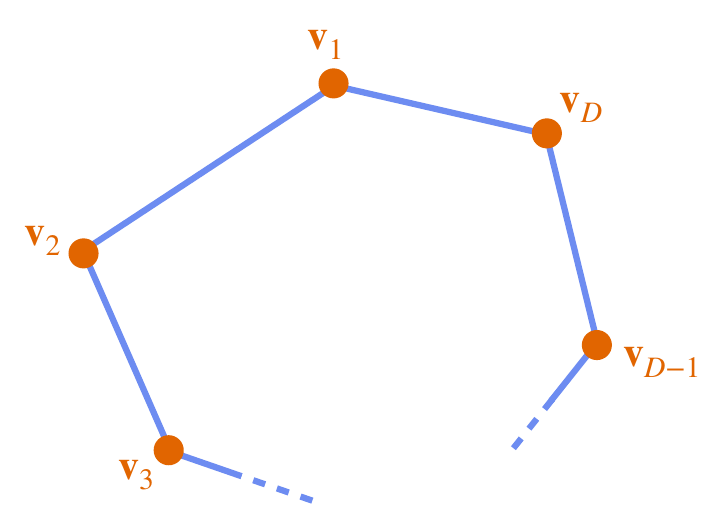}
\captionof{figure}{The toric diagram of K\"ahler cone $X_6$, with toric fan vectors $\{\v_I\}$.}\label{fig: toric}
\end{minipage}
\end{center}
Then, $X_6$ can be understood as a $T^3$ fibration over a polyhedral cone $\C\subset \R^3$. This cone is in turn specified entirely in terms of a set of primitive \textit{toric fan vectors} $\v_I\in \Z^3$, $I=1,\dots,D$, for some $D\ge 3$. The Gorenstein condition means we can write $\v_I = (v_I^1,v_I^2,1)$, and thus the ends of all fan vectors lie in a plane. Furthermore, the convex hull in $\R^2$ of the endpoints $(v_I^1,v_I^2)\in \Z^2\subset \R^2$ is such that all points $(v_I^1,v_I^2)$ lie on the boundary; that is to say, the polygon formed by joining these endpoints is convex\footnote{Implicit here is that all points in $\Z^2$ that lie on the boundary of the toric diagram are the projection $(v_I^1,v_I^2)$ to $\Z^2$ of some fan vector $(v_I^1,v_I^2,1)$. Said another way, the edge in $\R^2$ connecting any two consecutive fan vectors $(v_I^1,v_I^2)$ and  $(v_1^{I+1},v_2^{I+1})$ does not pass through any point in $\Z^2$. For simplicity, we specialise to toric diagrams such that each fan vector is a true corner. This is precisely the restriction that for each $I=1,\dots,D$, the integer $\v_{I-1}\cdot (\v_I \times \v_{I+1})$ is strictly positive (while in general it is only non-negative). In the general case, one has to deal with certain shrinking 3-cycles in the geometry, which in turn complicates the giant graviton expansion somewhat \cite{Fujiwara:2023bdc}.  \label{footnote: coprime}}.
This polygon defines the \textit{toric diagram}; see Figure \ref{fig: toric}. Note that we order the fan vectors in an anti-clockwise fashion. Then, the cone $\C$ is defined by
\begin{align}
  \C=\{\u\in \R^3 \,\,:\,\, \u \cdot \v_I \ge 0,\,\, I=1,\dots,D\}\ .
\end{align}
The $T^3$ fibration is non-degenerate in the interior of the cone, but a particular circle degenerates on each bounding facet $\{\u\cdot \v_I = 0\}$.

\para
Then, the five-manifold $Y_5 = X_6 |_{r=1}$, known as the \textit{link} of the cone, is a \textit{toric Sasakian} manifold. Like $X_6$, $Y_5$ can be thought of as a $T^3$ fibration, now over a compact, convex two-dimensional polytope $\P$. This is found as a certain slice of the cone $\C$,
\begin{align}
  \P = \C \cap \H(\b)\ ,
\end{align}
where $\H(\b)$ is the \textit{Reeb hyperplane}
\begin{align}
  \H(\b):= \{\u\in \R^3 \,\, : \,\, \u\cdot \b = \tfrac{1}{2}\}\ .
\end{align}
A key result of \cite{Martelli:2005tp} is that for a specific choice of the Reeb vector $\b$, which is determined by an extremisation principle subject to the constraint $b^3 = 3$, $Y_5$ is also Einstein. This then defines a \textit{Sasaki-Einstein} manifold $Y_5 = \text{SE}_5$.

\para
The simplest example of this setup is simply $X_6=\mathbb{C}^3$, which has $D=3$. The link is then a 5-sphere, $\text{SE}_5=S^5$, which is indeed a $T^3$ fibration over a triangle $\P$. Each edge corresponds to a $T^2$ fibration over an interval, which is just an $S^3$, while each corner corresponds to an $S^1$.

\para
Now, given some choice of fan vectors $\v_I$ and Reeb vector $\b$, there is a further way in which we can generalise the geometry $Y_5$, by varying its K\"ahler class. This amounts to introducing a broader set of geometries, parameterised not only by $\v_I$ and $\b$, but also a set of $D$ real parameters $\lambda_I$. Such a geometry is not generically Sasakian; we revert to the previous Sasakian geometry by setting $\lambda_I = -1/2b^3$ for each $I=1,\dots,D$.

\para
To summarise, we can define a set of five-dimensional geometries $Y_5$ depending on fan vectors $\v_I\in \R^3$, Reeb vector $\b\in \R^3$, and K\"ahler moduli $\lambda_I\in \R$. By fixing the $\lambda_I$ to a particular value, $Y_5$ becomes Sasakian. By further fixing $\b$ to a particular value, $Y_5$ becomes Sasaki-Einstein, $Y_5 = \text{SE}_5$.

\para
Now, one can define and compute a \textit{volume} associated to the generic geometry $Y_5$, dependent on $\lambda_I, \b$ as well as the $\v_I$, which is referred to as the \textit{master volume} \cite{Gauntlett:2018dpc} due to its utility in computing a broad range of supergravity quantities. Explicitly, it is given by
\begin{align}
  &\V(\b,\lambda_I) \nn\\
  &\quad = \frac{(2\pi)^3}{2} \sum_{I=1}^D \left(\frac{\lambda_I \left(\lambda_I\,\b \cdot (\v_{I+1}\times \v_{I-1})+ \lambda_{I+1}\,\b \cdot (\v_{I-1}\times \v_{I}) + \lambda_{I-1} \,\b \cdot (\v_{I}\times \v_{I+1})\right)}{\b \cdot (\v_{I-1}\times \v_{I})\,\,\b \cdot (\v_{I}\times \v_{I+1})}\right)\ .
\label{eq: AdS5 master volume}
\end{align}
Before we move on, note that the K\"ahler class of $Y_5$ in fact depends only on $(D-2)$ of the parameters $\lambda_I$, and as such, $\V(\b,\lambda_I)$ enjoys a two-dimensional shift symmetry. In detail, we have
\begin{align}
  \V(\b,\lambda_I) = \V(\b,\lambda_I')\ ,
  \label{eq: master volume symmetry}
\end{align}
for any two sets of K\"ahler moduli related by
\begin{align}
  \lambda_I' = \lambda_I + \u \cdot \left(b^3 \v_I - \b \right)\ ,
\end{align}
for some $\u\in \R^3$. At first this looks like three degrees of symmetry, but notice that $v_I^3 = 1$, and hence, $\u \cdot \left(b^3 \v_I - \b \right)$ does not depend on $u^3$.

\subsection{The giant graviton index}\label{subsec: AdS5 GGI}

We now pick some toric SE$_5$, which as just discussed, is entirely specified by a set of fan vectors $\v_I$.
Then, there is a recipe to read off the four-dimensional $\N=1$ SCFT dual to AdS$_5\times \text{SE}_5$ \cite{Hanany:2005ve,Franco:2005sm,Franco:2005rj,Feng:2005gw}. It is this theory that we now study.

\para
The basic object of study is the \textit{superconformal index}, which we can define as follows. In addition to the scaling dimension $\Delta$ and spin $J_i$, $i=1,2$, Cartan generators for the conformal group $SO(2,4)$, states are organised by their charges $Q_I, I=1,\dots, D$ under a $U(1)^D$ global symmetry\footnote{A particularly combination of these global rotations, determined by $a$-maximisation \cite{Intriligator:2003jj}, forms the $U(1)_R$ R-symmetry of the $\N=1$ superconformal group.}. One can always choose a convention for these global charges such that the Poincar\'e supercharge $\Q$ carries charges 
\begin{align}
  (+\tfrac{1}{2},-\tfrac{1}{2},+\tfrac{1}{2}) \qquad \text{under} \,\, (\Delta,J_i,Q_I)\ .
\end{align}
Then, the superconformal index can be defined as a trace over the Hilbert space of the theory on $S^3$, 
\begin{align}
  \I(\omega_i,\Delta_I) = \text{Tr} \left[e^{-\omega_i J_i - \Delta_I Q_I}\right]\ ,
\end{align}
and is a function of some generally complex valued chemical potentials $(\omega_i, \Delta_I)$. In particular, the requirement that this does indeed define an index for $\Q$ is precisely that these chemical potentials satisfy
\begin{align}
  \omega_1 + \omega_2 - \sum_{I=1}^D \Delta_I = 2\pi i \quad  (\text{mod }4\pi i)\ ,
  \label{eq: linear relation}
\end{align}
which ensures that all states above the BPS bound $\{\Q,\Q^\dagger\}\ge 0$ cancel in pairs. The index $\I(\omega_i,\Delta_I)$ can equivalently be regarded as a Euclidean partition function of the theory on $S^1\times S^3$, with twisted boundary conditions on the $S^1$.
\para
Comparison with (semi-)classical gravity in the bulk necessitates that we understand the large $N$ asymptotics of $\I(\omega_i, \Delta_I)$. In a suitable regime of chemical potential space, one finds the following asymptotic \cite{Choi:2023tiq,Amariti:2019mgp,Lanir:2019abx,Benini:2020gjh,Cabo-Bizet:2018ehj,Cabo-Bizet:2019osg,Cabo-Bizet:2020nkr}
\begin{align}
  \log \I(\omega_i,\Delta_I) \sim  \frac{ N^2}{12\omega_1\omega_2}\sum_{I,J,K}C_{I,J,K}\Delta_I \Delta_J \Delta_K\ , \qquad C_{I,J,K}:= |\v_I \cdot (\v_J \times \v_K)|\ ,
  \label{eq: total index asymptotic}
\end{align}
which is indeed consistent with all known relevant supergravity results. In particular, the right-hand-side of this expression should coincide with minus the Euclidean on-shell action of the dominant black hole saddle in the background as fixed by the values of $\{\omega_i, \Delta_I\}$. Furthermore, by Legendre transforming\footnote{This is not quite simply a change of ensemble, which by basic thermodynamics must produce the entropy, because we can only vary chemical potentials subject to the constraint (\ref{eq: linear relation}). Rather, this Legendre transform takes us to a mixed ensemble \cite{Mouland:2023gcp}, which in a great range of scenarios nonetheless reproduces the microcanonical entropy \cite{Cassani:2019mms}. This is closely related to the so-called `non-linear constraint' on black hole charges \cite{Mouland:2023gcp}.} in chemical potential space, one lands on the Bekenstein-Hawking entropy of this black hole; for this reason, this object is sometimes called the \textit{entropy function} of the black hole. We will have much more to say about this object in Section \ref{subsec: AdS5 BH entropy}.

\para
Our interest is in an interesting proposed reformulation of $\I(\omega_i,\Delta_I)$ in terms of its \textit{giant graviton expansion} \cite{Imamura:2021ytr,Gaiotto:2021xce,Murthy:2022ien,Bourdier:2015sga,Bourdier:2015wda,Arai:2019aou,Arai:2020uwd}. The idea is simple: we should be able to reproduce the index $\I(\omega_i,\Delta_I)$ by summing over BPS states in the bulk, which in addition to supergravity/string modes also include supersymmetric D-brane configurations. How this works in this AdS$_5\times \text{SE}_5$ case has been discussed in \cite{Arai:2019aou,Fujiwara:2023bdc,Kim:2024ucf,Hatsuda:2024lcc}, which we now briefly review. There is a supersymmetric 3-cycle $S_I$ in $\text{SE}_5$ associated to each fan vector $\v_I$, which is topologically a lens space,
\begin{align}
  S_I \cong S^3/\Z_{k_I},\qquad k_I = \v_{I-1}\cdot (\v_I\times \v_{I+1}) >0\ ,
\end{align}
where we treat the indices cyclically, so that $\v_{D+1}=\v_1$.
Note that there are precisely $D$ non-empty intersections of these 3-cycles, which take the form
\begin{align}
  S_I \cap S_{I+1} \cong S^1\ .
\end{align}
A supersymmetric 3-cycle in SE$_5$ is labelled by $D$ non-negative integers $(n_1,\dots,n_D)$, signifying that it wraps $S_1$ $n_1$ times, $S_2$ $n_2$ times, and so on. Then,  $\I(\omega_i, \Delta_I)$ receives a contribution from a D3-brane which wraps the $(n_1,\dots, n_D)$ 3-cycle. Summing these contributions, we arrive at the expansion
\begin{align}
  \I(\omega_i,\Delta_I) = \I_\infty (\omega_i,\Delta_I) \sum_{n_I\in \Z_{\ge0}} e^{-N n_I \Delta_I} \,\I_{(n_1,\dots,n_D)} (\omega_i,\Delta_I)\ ,
\end{align}
where $\I_\infty (\omega_i,\Delta_I)$ captures the graviton contribution. Meanwhile, the \textit{giant graviton index} $\I_{(n_1,\dots,n_D)} (\omega_i,\Delta_I)$ is an index that counts BPS ground states of the D3-brane configuration, and can be understood as follows.

\para
The giant graviton index is the superconformal index of a $U(n_1)\times \dots \times U(n_D)$ gauge theory on the D3-branes configuration. This is a 4d-2d coupled system, made up of four-dimensional modes associated to each 3-cycle $S_I$, coupled to two-dimensional modes associated to each intersection $S_I \cap S_{I+1}$. Schematically, it is given by the matrix integral
\begin{align}
  \I_{(n_1,\dots,n_D)}(\omega_i , \Delta_I) =\int \left(\prod_{I=1}^D d\,^{n_I}u_I \right) \left(\prod_{I=1}^D F_I^{(4d)}(\omega_i, \Delta_I; u_I)\right)\left(\prod_{I=1}^D F_{I,I+1}^{(2d)}(\omega_i, \Delta_I; u_I, u_{I+1})\right)\ ,
  \label{eq: GG index schematic}
\end{align}
where $u_I$ denotes a set of $n_I$ chemical potentials for the $U(n_I)$ gauge symmetry, normalised such that $g=e^{2\pi i\, \text{diag}((u_I)_1,\dots,(u_I)_{n_I})}$ parameterises the $U(1)^{n_I}\subset U(n_I)$ Cartan subgroup. Note that in our conventions, the relevant Haar measure is packaged within the $F_I^{(4d)}$.

\para
We now study the large $n_I$ asymptotics of $\I_{(n_1,\dots,n_D)}$. Our strategy can be viewed as an extension of the `parallelogram ansatz' approach of \cite{Choi:2021rxi}\footnote{See also \cite{Choi:2023tiq,Kim:2024ucf}. In particular, the key extension considered in this work is to a 4d-2d coupled system, which was touched upon in \cite{Kim:2024ucf}.}. We outline the basic steps here, relegating full details to Appendix \ref{app: asymptotics}.

\para
Each four-dimensional contribution $F_I^{(4d)}$ is given by a certain ratio of elliptic gamma functions. The idea then is to use the modular properties of these functions to write
\begin{align}
  F_I^{(4d)}(\omega_i, \Delta_I; u_I) = \modker^{(4d)}_I(\omega_i,\Delta_I) \tilde{F}_I^{(4d)}(\omega_i, \Delta_I; u_I)\ ,
\label{eq: 4d modular}
\end{align}
where the modular kernel $\modker^{(4d)}_I$ is independent of the integration variables $u_I$. Similarly, each two-dimensional contribution $F_{I,I+1}^{(2d)}$ is given by a ratio of $q$-theta functions, whose modular properties allow us to write
\begin{align}
  F_{I,I+1}^{(2d)}(\omega_i, \Delta_I; u_I, u_{I+1}) = \modker^{(2d)}_{I,I+1}(\omega_i,\Delta_I) \tilde{F}_I^{(2d)}(\omega_i, \Delta_I; u_I, u_{I+1})\ ,
\label{eq: 2d modular}
\end{align}
where again, the modular kernel $\modker^{(2d)}_{I,I+1}$ is independent of the integration variables. In all, we thus have the recasting
\begin{align}
  \I_{(n_1,\dots,n_D)}(\omega_i , \Delta_I) = \left(\prod_{I=1}^D \modker_I^{(4d)}(\omega_i, \Delta_I) \modker_{I,I+1}^{(2d)}(\omega_i, \Delta_I) \right) \tilde{\I}_{(n_1,\dots,n_D)}(\omega_i , \Delta_I)\ ,
\end{align}
where importantly, the modular pre-factor exhibits exponential growth at large $n_I$. In particular, we shall see that $\log \modker_I^{(4d)}$ and $\log\modker_{I,I+1}^{(2d)}$ each scale quadratically in the $n_I$ as $n_I\to \infty$.

\para
The name of the game, then, is to engineer the modular transformations (\ref{eq: 4d modular}) and (\ref{eq: 2d modular}) such that the remaining integral 
\begin{align}
  \tilde{\I}_{(n_1,\dots,n_D)}(\omega_i , \Delta_I) =\int \left(\prod_{I=1}^D d\,^{n_I}u_I \right) \left(\prod_{I=1}^D \tilde{F}_I^{(4d)}(\omega_i, \Delta_I; u_I)\right)\left(\prod_{I=1}^D \tilde{F}_{I,I+1}^{(2d)}(\omega_i, \Delta_I; u_I, u_{I+1})\right)\ ,
\label{eq: remaining integral}
\end{align}
provides a contribution to the asymptotics of $\log  \I_{(n_1,\dots,n_D)}$ at large $n_I$ that is subleading to the contribution of the modular kernels $\modker^{(4d)}_I$ and $\modker^{(2d)}_{I,I+1}$. Demonstrating such behaviour for $ \tilde{\I}_{(n_1,\dots,n_D)}$ is the crux of this approach, and proceeds as follows. First, in the case that some orbifold number $k_I$ is greater than $1$, the corresponding lens space partition function $F^{(4d)}_I$ and hence its modular transform $\tilde{F}^{(4d)}_I$ decomposes into a sum of $k_I^{n_I}$ terms, corresponding to the different choices of gauge holonomy on the lens space $S_I = S^3/\Z_{k_I}$. Hence, $ \tilde{\I}_{(n_1,\dots,n_D)}$ decomposes into a sum of $k_1^{n_1}k_2^{n_2}\dots k_D^{N_D}$ integrals. We can then tackle each of these in isolation. In particular, we observe that each corresponding integrand possesses a very particular periodicity structure, which allows one to exhibit a large $n_I$ saddle-point. Under the assumption that this saddle dominates the integral, we find that under certain conditions that we discuss below, each integral is $\mathcal{O}(1)$, and thus after summing back up over holonomy sectors, we find 
\begin{align}
   \log \tilde{\I}_{(n_1,\dots,n_D)} (\omega_i , \Delta_I) \sim \sum_I n_I \log k_I\ ,
\end{align}
as $n_I\to \infty$. In particular, this is subleading to the $\mathcal{O}(n_I^2)$ growth of the modular kernels.

\para
Thus, the leading asymptotic for $\I_{(n_1,\dots,n_D)}(\omega_i , \Delta_I)$ is given by
\begin{align}
  \log \I_{(n_1,\dots,n_D)}(\omega_i , \Delta_I) \sim \sum_{I=1}^D \log \modker^{(4d)}_I (\omega_i , \Delta_I) + \sum_{I=1}^D \log \modker^{(2d)}_{I,I+1} (\omega_i , \Delta_I)\ .
\label{eq: total modker}
\end{align}
We now provide the explicit expressions for each contribution here, and interpret them in terms of the relevant four-dimensional and two-dimensional SCFTs. To do so, it is useful to define
\begin{align}
  \beps = \sum_{I=1}^D \Delta_I \v_I\ .
\end{align}
Then, the various contributions are given as follows.

\begin{itemize}
  \item \textbf{Four-dimensional contributions}\\[0.4em]
  We have
  \begin{align}
  \log \modker^{(4d)}_I (\omega_i , \Delta_I) =  \frac{n_I^2 \,\omega_1\omega_2 }{2}  \left(\frac{\beps \cdot (\v_{I-1}\times \v_{I+1})}{\beps \cdot (\v_{I-1}\times \v_{I})\,\,\beps \cdot (\v_{I}\times \v_{I+1})}\right)\ .
  \label{eq: 4d mod ker}
  \end{align}
Essentially by construction, this coincides with the $S^1\times S_I = S^1\times (S^3/\Z_{k_I})$ free energy of the $U(n_I)$ $\N=4$ super-Yang-Mills theory living on $n_I$ D3-branes. Explicitly, 
  \begin{align}
  \log \modker^{(4d)}_I (\omega_i , \Delta_I) =  \frac{n_I^2}{2k_I} \frac{\Delta_1^{(I)}\Delta_2^{(I)}\Delta_3^{(I)}}{\omega_1^{(I)}\omega_2^{(I)}}\ .
  \label{eq: 4d free energy}
\end{align}
The chemical potentials appearing here parameterise twisted boundary conditions around the `thermal' circle, with $\omega_{1.2}^{(I)}$ corresponding to twists in the $SO(4)$ worldvolume rotation symmetry, and $\Delta_{1,2,3}^{(I)}$ to twists in the $SO(6)$ R-symmetry. Supersymmetry requires they satisfy
\begin{align}
  \omega_1^{(I)} + \omega_2^{(I)} - \Delta_1^{(I)}- \Delta_2^{(I)}- \Delta_3^{(I)} = 2\pi i \quad (\text{mod }4\pi i)\ .
  \label{eq: index constraint again}
\end{align}
All of these potentials can be deduced from the form of the metric on AdS$_5\times \text{SE}_5$ with chemical potentials $(\omega_i,\Delta_I)$. In particular, $\omega^{(I)}_{1,2}$ are read off from the induced metric on $S^1\times S_I$, while $\Delta_{1,2,3}^{(I)}$ correspond to metric components in the transverse directions. They are thus in turn fixed in terms of the toric data defining SE$_5$. We find
\begin{align}
  &\omega^{(I)}_1 = \frac{1}{k_I} \beps\cdot(\v_{I-1}\times \v_{I})\ ,\qquad \omega^{(I)}_2 = \frac{1}{k_I} \beps\cdot(\v_{I}\times \v_{I+1})\ ,\nn\\
    &\Delta^{(I)}_1 = \omega_1\ ,\qquad \Delta^{(I)}_2 = \omega_2\ ,\qquad \Delta^{(I)}_3 = \frac{1}{k_I} \beps\cdot(\v_{I-1}\times \v_{I+1})\ , 
\label{eq: 4d chem pots}
\end{align}
which do indeed satisfy (\ref{eq: index constraint again}), by virtue of the identity
\begin{align}
  \v_J \cdot \left(\v_{I-1}\times \v_I + \v_I\times \v_{I+1} + \v_{I+1}\times \v_{I-1}\right) =  k_I\ ,
\end{align}
for each $J=1,\dots,D$. Plugging these into (\ref{eq: 4d free energy}), we indeed recover the result (\ref{eq: 4d mod ker}).
  
  \item \textbf{Two-dimensional contributions}\\[0.4em]
  We have
  \begin{align}
  \log \modker^{(2d)}_{I,I+1}(\omega_i, \Delta_I) =  -\frac{n_I n_{I+1}\omega_1 \omega_2}{\beps\cdot(\v_{I}\times \v_{I+1})}\ .
  \label{eq: 2d result main text}
\end{align}
This can then be interpreted as the $S^1\times S^1$ free energy of the two-dimensional SCFT that describes the open string modes localised to the intersection $S_I \cap S_{I+1}\cong S^1$,
\begin{align}
  \log \modker^{(2d)}_{I,I+1}(\omega_i, \Delta_I) = -n_I n_{I+1} \frac{\Delta_1^{(I,I+1)}\Delta_2^{(I,I+1)}}{\omega^{(I,I+1)}}\ ,
\end{align}
where we identify
\begin{align}
  \Delta_1^{(I,I+1)} = \omega_1,\quad \Delta_2^{(I,I+1)} = \omega_2,\quad \omega^{(I,I+1)} =  \beps\cdot(\v_{I}\times \v_{I+1})\ .
\end{align}

\end{itemize}
We thus ultimately arrive at the compact asymptotic
\begin{align}
  \log \I_{(n_1,\dots,n_D)}(\omega_i,\Delta_I) &\sim  \frac{\omega_1 \omega_2}{2} \sum_{I=1}^D \left(\frac{n_I^2\,\beps \cdot (\v_{I-1}\times \v_{I+1})}{\beps \cdot (\v_{I-1}\times \v_{I})\,\,\beps \cdot (\v_{I}\times \v_{I+1})} - \frac{2n_I n_{I+1}}{\beps\cdot(\v_{I}\times \v_{I+1})}\right)		\nn\\
  &\hspace{-30mm}=  -\frac{\omega_1 \omega_2}{2} \sum_{I=1}^D \left(\frac{n_I \left(n_I\,\beps \cdot (\v_{I+1}\times \v_{I-1})+ n_{I+1}\,\beps \cdot (\v_{I-1}\times \v_{I}) + n_{I-1} \,\beps \cdot (\v_{I}\times \v_{I+1})\right)}{\beps \cdot (\v_{I-1}\times \v_{I})\,\,\beps \cdot (\v_{I}\times \v_{I+1})}\right)	\ ,
  \label{eq: final GG index asymptotic}
\end{align}
where recall the indices on $\v_I$ are cyclic, so that $\v_{I+D}=\v_I$.

\para
Then, by an immediate comparison with (\ref{eq: AdS5 master volume}), we establish the main proposal of this paper:
\begin{align}
  \boxed{\phantom{\bigg(}\log \I_{(n_1,\dots,n_D)}(\omega_i,\Delta_I) \sim -\frac{\omega_1\omega_2}{(2\pi)^3} \,\V(\b,\lambda_I )\phantom{\bigg)}}
\label{eq: AdS5 result}
\end{align}
where we identify Reeb vector and K\"ahler moduli as
\begin{align}
  \b= \beps = \sum_I \Delta_I \v_I,\qquad \lambda_I = n_I\ .
\end{align}
Before moving on, let us make a few remarks.

\begin{itemize}
  \item The key computational step in deriving the asymptotic (\ref{eq: AdS5 result}) is the verification that the modular-transformed integral $\log\tilde{\I}_{(n_1,\dots,n_D)}$ is subleading to the $\mathcal{O}(n_I^2)$ growth of the modular kernels (\ref{eq: total modker}) as $n_I\to \infty$. The validity of this statement, as detailed in Appendix \ref{app: asymptotics}, is subtle. It requires that the exhibited saddle-point configuration of the matrix integral $\tilde{\I}_{(n_1,\dots,n_D)}$ avoids certain branch points, which ultimately imposes a set of constraints on the chemical potentials $(\omega_i, \Delta_I)$. We thus carve out a region, say Region A, of chemical potential space $(\omega_i, \Delta_I)$, in which our asymptotic is valid.
  \item There is a way in which we can expand our result to a broader region of chemical potential space, which relies on a simplification of the giant graviton expansion (\ref{eq: GG expn intro}) proposed in \cite{Fujiwara:2023bdc} known as the `simple sum' reduction. This says that the sum in (\ref{eq: GG expn intro}) is unchanged if we choose to sum only over giant graviton configurations with two adjacent nodes turned off, say $n_{D-1}=n_D=0$. The virtue of this formulation is that the constraints on chemical potentials such that $\log\tilde{\I}_{(n_1,\dots,n_{D-2},0,0)}$ is subleading (versus quadratic in the $n_I$) are weaker than for $\log\tilde{\I}_{(n_1,\dots,n_{D})}$. We can thus define a Region B of $(\omega_i,\Delta_I)$ space, containing Region A, in which the corresponding asymptotic (\ref{eq: AdS5 result}) for $\I_{(n_1,\dots,n_{D-2},0,0)}$ is valid. 
  \item The determination of Regions A and B for each choice of SE$_5$ is rather involved, and is left to future work. For now, we provide in Appendix \ref{app: asymptotics} a detailed analysis of Region B in the case of\footnote{Note that this region is, essentially by construction, qualitatively identical to the regime of validity of the asymptotic of the $\N=4$ super-Yang-Mills index itself as exhibited in \cite{Choi:2021rxi}. Thus, our analysis can be repurposed in this case.} SE$_5=S^5$.  
  \item It was recognised in \cite{Choi:2021rxi} that different asymptotics of matrix integrals like $\tilde{\I}_{(n_1,\dots,n_D)}$ arise as we move between different regions of chemical potential space, and that in the case of the $\N=4$ super-Yang-Mills superconformal index, these alternative asymptotics could be related to the alternative saddles of \cite{Aharony:2021zkr}. It would be interesting to pursue a similar analysis for the giant graviton index $\tilde{\I}_{(n_1,\dots,n_D)}$.
\end{itemize}

%

\subsection{Black hole entropy}\label{subsec: AdS5 BH entropy}

We finally discuss the relationship between the large $n_I$ asymptotic (\ref{eq: final GG index asymptotic}) of the giant graviton index, and the large $N$ asymptotic (\ref{eq: total index asymptotic}) of the index as a whole, which as mentioned above coincides with the entropy function of the corresponding BPS black hole. 

\para
We have
\begin{align}
  &\frac{\I(\omega_i,\Delta_I)}{\I_\infty(\omega_i, \Delta_I) } \sim  \sum_{n_I\in \Z_{\ge 0}} e^{- \A(n_I)} \ ,
  \label{eq: expansion ready for saddle-point} 
  \end{align}
  where
  \begin{align}
  \A(n_I)=  N \sum_I n_I \Delta_I - \frac{\omega_1 \omega_2}{2} \sum_I \left(\frac{n_I^2\,\beps \cdot (\v_{I-1}\times \v_{I+1})}{\beps \cdot (\v_{I-1}\times \v_{I})\,\,\beps \cdot (\v_{I}\times \v_{I+1})} - \frac{2n_I n_{I+1}}{\beps\cdot(\v_{I}\times \v_{I+1})}\right) \ ,
  \label{eq: A def}
\end{align}
whose dependence on $\omega_i, \Delta_I$ we suppress in our notation. The aim then is to determine the dominant giant graviton configuration $\{n_I\}$ in this sum at large $N$. We do so by first appealing to a continuum approximation, so as to treat the $n_I$ as continuous variables,
\begin{align}
  &\frac{\I(\omega_i,\Delta_I)}{\I_\infty(\omega_i, \Delta_I) } \sim \int_0^\infty \prod_I dn_I \,\, e^{- \A(n_I)} \ ,
  \label{eq: continuum approx} 
  \end{align}
 before applying a saddle-point approximation in $n_I$ space. We will later check the self-consistency of this approximation, once we have studied the saddle-point structure of the integral.

\para
The special case SE$_5 = S^5$ of this calculation was explored in \cite{Beccaria:2023hip,Choi:2022ovw}, which was then extended to some specific other choices of SE$_5$ in \cite{Kim:2024ucf}; our analysis can be seen as an extension of these results to generic SE$_5$. However we also offer an alternative perspective on a key first step in these existing analyses\footnote{Similar observations to ours have been made in the special case of SE$_5=S^5$ in \cite{Beccaria:2023hip}.}. Namely, it is noticed in these works that the function $\A(n_I)$ exhibits two flat directions, rendering a saddle-point analysis potentially problematic. This issue is often then circumvented by proposing that the sum (\ref{eq: expansion ready for saddle-point}) can in fact first be truncated to the `simple sum', as discussed above, in which one sums over only $(D-2)$ of the $D$ wrapping numbers. On this reduced space of configurations, a unique saddle-point is found, and the relevant entropy function recovered. 

\para
Our perspective on and treatment of these flat directions is rather different. We already know that the second term in $\A(n_I)$ exhibits two degrees of gauge symmetry in the $n_I$, which are nothing other than the geometric gauge symmetries (\ref{eq: master volume symmetry}) of the master volume. It will turn out that these are also symmetries of the first term, and thus we provide a concrete geometric origin for the two flat directions in (\ref{eq: continuum approx}). We argue however that while highly non-trivial cancellations in the index may occur along these flat directions, giving rise to the `simple-sum' reformulation, this mechanism is not necessary for the saddle-point analysis to go through. This is precisely because while the saddle-points of the integral (\ref{eq: continuum approx}) do indeed form a two-parameter continuous family, given by a single gauge orbit, the volume of this gauge orbit is in fact finite. We will compute its volume, which provides a $\log N$ term in the large $N$ expansion of $\log \I(\omega_i, \Delta_I)$. 

\para
Conversely, it is clear that the equality in magnitude of the saddle-point contributions from giant graviton configurations lying on the same gauge orbit is very much supportive of the proposed `simple sum' reduction. While at leading order we don't see these cancellations, consistency with this picture suggests that by studying higher order corrections both to the large $n_I$ asymptotics of $\I_{(n_1,\dots,n_D)}$ and the saddle-point evaluation of the integral in (\ref{eq: continuum approx}), one should find relative signs between contributions on the same gauge orbit, thus signalling significant cancellations. We leave such a study to future work.

\para
Our first step is to exhibit the two flat directions of $\A$ more explicitly. We have
\begin{align}
  \A(n_I') = \A(n_I)\ ,
  \label{eq: invariance equation}
\end{align}
for any configurations $\{n_I\},\{n_I'\}$ related by
\begin{align}
  n_I' = n_I +\sum_{J=1}^D \Delta_J \,\u \cdot (\v_I -\v_J),\qquad \text{for any }\u  \in \R^3\ .
\end{align}
At first blush this looks like three degrees of invariance, but note that $(0,0,1)\cdot (\v_I - \v_J)=0$ and thus $\left(n_I' - n_I\right)$ depends only on the first two components of $\u$. We therefore have two degrees of invariance. This is of course just the symmetry (\ref{eq: master volume symmetry}) of the master volume under shifts of K\"ahler moduli, which turns out to also be a symmetry of the first term in (\ref{eq: A def}), since
\begin{align}
  \sum_I (n_I' - n_I)\Delta_I = \sum_{I,J} \Delta_I \Delta_J \u \cdot (\v_I -\v_J) = 0\ .
\end{align}
Thus, we can reduce the integral in (\ref{eq: continuum approx}) to one over only $(D-2)$ variables. To do so, we introduce new integration variables $(m_A,\sigma_1,\sigma_2)$, where $A=1,\dots,D-2$, related to the $n_I$ by
 \begin{align}
  n_I =\left\{\begin{aligned}
	\,\,m_I \,\,+ &\,\,\sum_J \Delta_J \,\u \cdot (\v_I - \v_J),\quad  I=1,\dots, D-2		\\
	 &\,\,\sum_J \Delta_J \,\u \cdot (\v_I - \v_J),\quad  I=D-1,D
\end{aligned}\right.
 \qquad \text{where } \u = \begin{pmatrix}
  	\sigma_1\\\sigma_2\\ 0
  \end{pmatrix}\ .
\end{align}
We then have
\begin{align}
  \int_0^\infty \prod_I dn_I\, e^{- \mathcal{A}(n_I)} =  \int_0^\infty \prod_A dm_A \,\, V(m_A)\,\, e^{-\B(m_A)}\ ,
\end{align}
where
  \begin{align}
  \B(m_A)=  N \sum_{A=1}^{D-2} m_A \Delta_A - \frac{\omega_1 \omega_2}{2} \sum_{A=1}^{D-2} \left(\frac{m_A^2\,\beps \cdot (\v_{A-1}\times \v_{A+1})}{\beps \cdot (\v_{A-1}\times \v_{A})\,\,\beps \cdot (\v_{A}\times \v_{A+1})} - \frac{2m_A m_{A+1}}{\beps\cdot(\v_{A}\times \v_{A+1})}\right) \ .
\end{align}
Note that the final term in the second sum has only a single term, and recall the indices on the $\v_I$ are cyclic, so that $\v_{0} = \v_D$. Meanwhile, $V(m_A)$ is the volume of the gauge orbit of the configuration $n_I = (m_1,\dots,m_{D-2},0,0)$, given by
\begin{align}
  V(m_A) = J\iint_{P(m_A)\subset \R^2} d\sigma_1 d\sigma_2 = J\times \text{Area}[P(m_A)] \ ,
  \label{eq: orbit volume def}
\end{align}
where $J$ is the Jacobian arising from the change of variables (which is independent of $m_A$ and $\sigma_{1,2}$ since the transformation was linear). All non-triviality comes in the integration region, which is a polygon $P(m_A)\subset \R^2$ defined, for given $\{m_A\} $, by the points in the $(\sigma_1,\sigma_2)$ plane such that $n_I(m_A,\sigma_1,\sigma_2)\ge 0$. A precise definition of $P(m_A)$ and a computation of its area can be found in Appendix \ref{app: gauge orbit},  where we find the result (\ref{eq: orbit volume result}) for $V(m_A)$. In particular, it is polynomial in the $m_A$, and thus is irrelevant when it comes to determining the saddle-point value of the $m_A$.

\para
Applying a saddle-point approximation, we thus have
\begin{align}
  \int_0^\infty \prod_I dn_I\, e^{- \mathcal{A}(n_I)} \sim V(m_A^*) \,e^{-\B(m_A^*)}\ .
\end{align}
The saddle-point equations are\footnote{Note that these same equations, and their solutions below, have appeared in a conceptually distinct but computationally similar context \cite{Hosseini:2019use}.}
\begin{align}
  \frac{\partial \B}{\partial m_A} (m_A^*) = 0\ .
\end{align}
This is a set of $(D-2)$ linearly-independent linear equations for the $(D-2)$ variables $m_A$, which thus have a unique solution. It is
\begin{align}
  m^*_A = -\frac{N}{\omega_1 \omega_2} (\beps \times \v_A) \cdot \left(\sum_{I=A+1}^{D-1} \Delta_I \v_I\right)\ .
\label{eq: saddle-point}
\end{align}
We find then that
\begin{align}
  \B(m_A^*) = -\frac{ N^2}{12\omega_1\omega_2}\sum_{I,J,K}C_{I,J,K}\Delta_I \Delta_J \Delta_K\ ,
\end{align}
with $C_{I,J,K}$ as defined in (\ref{eq: total index asymptotic}). Note that the self-consistency of this saddle-point approximation is ensured if we take $N\gg 1$. Thus we have as $N\to \infty$,
\begin{align}
  \log \I(\omega_i, \Delta_I) &\sim \frac{ N^2}{12\omega_1\omega_2}\sum_{I,J,K}C_{I,J,K}\Delta_I \Delta_J \Delta_K  +  \log V(m_A^* ) + \dots\ .
\label{eq: BH asymptotic}
\end{align}
The leading order contribution thus matches precisely the expected black hole saddle as in (\ref{eq: total index asymptotic}).

\para
It is interesting that we also obtain a contribution to the $\log N$ correction to $\log \I$ due to the continuous degeneracy of the saddle-point. Explicitly, the saddle-point value $V(m_A^*)$ given in (\ref{eq: V at saddle}) gives rise to
\begin{align}
  \log V(m_A^*)= 2 \log N + \dots \ ,
\end{align}
up to terms that are finite at large $N$. The index will generically receive further $\log N$ terms both from 1-loop fluctuations both around the $n_I$ saddle-points, and from higher order in $n_I$ corrections the giant graviton index asymptotic (\ref{eq: final GG index asymptotic}), neither of which we study here. Indeed, by including these other contributions, one would expect to recover the known $\log N$ coefficient obtained via the Bethe-Ansatz approach \cite{GonzalezLezcano:2019nca,GonzalezLezcano:2020yeb}. 

\para
Let us finally note that if we \textit{do} choose to exploit the proposed `simple sum' reduction of the giant graviton expansion (\ref{eq: GG expn intro}), and thus sum only over the $\I_{(n_1,\dots,n_{D-2},0,0)}$, it is clear that we still arrive at the leading asymptotic (\ref{eq: BH asymptotic}). Now, the integral (\ref{eq: continuum approx}) is over just $(D-2)$ variables $n_A$, $A=1,\dots,D-2$, and the (now isolated) saddle-point lies at $n_A = m_A^*$. 

\subsection{Are BPS black holes giant gravitons?}

We have found that the black hole entropy function is recovered as the saddle-point value of the giant graviton expansion. So, do we have a recipe on our hands, to identify a particular configuration of giant gravitons as a particular BPS black hole? Almost, but not quite.

\para
It is instructive to work in the example SE$_5=S^5$, in which we have good control of the corresponding BPS black hole solutions. There is then a four real parameter\footnote{See \cite{Markeviciute:2018yal,Markeviciute:2016ivy} and more recently \cite{Kim:2023sig,Choi:2024xnv,Dias:2024edd} for a discussion on how the bulk realises the fifth direction in the BPS surface.} family of such rotating, charged BPS black holes. One can understand what this means in the canonical `ensemble', in which we hold fixed the four independent complex chemical potentials of the superconformal index. At generic points in $\mathbb{C}^4$, the large $N$ asymptotics of the index is not recovered in terms of the $e^{-(\text{On-shell action})}$ of a single BPS black hole, but rather can be understood as a sum over many such contributions. There is however a four real dimensional subspace $\mathcal{R}_\text{BH}\subset\mathbb{C}^4$, on which this sum is completely dominated by a single term, and we really can identify the index with $e^{-(\text{On-shell action})}$ for a specific BPS black hole.

\para
We can play a similar game with the giant graviton expansion. Let us in particular assume the simple sum reduction of the giant graviton expansion (\ref{eq: GG expn intro}), such that the sum runs only over $(n_1,n_2,n_3) = (0,0,n_3)$. The isolated saddle-point of the sum lives at
\begin{align}
  n_3 = n_3^* =  - \frac{\Delta_1 \Delta_2}{\omega_1 \omega_2}N\ ,
\end{align}
which, for generic points in the $\mathbb{C}^4$ chemical potential space is of course complex. What this is telling us is that at generic points in $\mathbb{C}^4$, the sum is not dominated by a single giant graviton configuration (which has $n_3$ real), but rather, its leading asymptotic is recovered as a sum over many such contributions. We can think of the relative sizes of these contributions forming a distribution, whose width grows with $|\text{Im}\, n_3^*|$. So in particular, there is a seven real dimensional subspace $\mathcal{R}_\text{GG}\subset \mathbb{C}^4$, where $- \frac{\Delta_1 \Delta_2}{\omega_1 \omega_2}$ is real and positive, on which this distribution becomes sharply peaked, and we really can understand the index as being dominated by a single giant graviton configuration.

\para
To summarise, at generic values of chemical potentials, the index at large $N$ can be thought of as sum of many BPS black hole configurations, or alternatively, a sum of many giant graviton configurations. There is a four real dimensional subspace $\mathcal{R}_\text{BH}\subset\mathbb{C}^4$ on which the former sum is dominated by a single term, and a seven real dimensional subspace $\mathcal{R}_\text{GG}\subset \mathbb{C}^4$ on which the latter sum is dominated by a single term. Thus, if $\mathcal{R}_\text{BH}\cap \mathcal{R}_\text{GG}$ is non-empty, then there exists a class of BPS black holes which can be identified, on the nose, as a specific configuration  $(0,0,n_3^*)$ of giant gravitons. So: is $\mathcal{R}_\text{BH}\cap \mathcal{R}_\text{GG}$ it non-empty?

\para
The answer is no: $\mathcal{R}_\text{BH}\cap \mathcal{R}_\text{GG}$ is empty. But in showing this, we'll see some interesting physics.

\para
Let's fix our chemical potentials $(\omega_1,\omega_2,\Delta_1,\Delta_2,\Delta_3)$ to live on $\mathcal{R}_\text{BH}$, with $\omega_1 + \omega_2 - \Delta_1 - \Delta_2 - \Delta_3 = -2\pi i$. A manifest parameterisation of this subspace can be determined by studying the relevant BPS black hole with unequal angular momenta and charges (subject to a constraint), found in \cite{Kunduri:2006ek}. Thankfully, we can take a more indirect route. The entropy $S$, angular momenta $J_1,J_2$, and charges $Q_1,Q_2,Q_3$ then satisfy the quantum statistical relation \cite{Gibbons:1976ue},
\begin{align}
  \frac{1}{2}N^2 \frac{\Delta_1 \Delta_2 \Delta_3}{\omega_1\omega_2} = S - \omega_i J_I - \Delta_I Q_I\ .
\end{align}
There are then certain linear combinations\footnote{These are precisely those linear combinations that commute with the supercharge with respect to which the index is defined.} of charges that can be written compactly in terms of the chemical potentials,
\begin{align}
  J_1 + Q_3 &= \frac{1}{2}N^2 \frac{\Delta_1 \Delta_2}{\omega_1 \omega_2}\left(\frac{\Delta_3}{\omega_1}-1\right)	\ ,	\nn\\
  J_2 + Q_3 &= \frac{1}{2}N^2 \frac{\Delta_1 \Delta_2}{\omega_1 \omega_2}\left(\frac{\Delta_3}{\omega_2}-1\right)	\ ,	\nn\\
  Q_3 - Q_1	&= \frac{1}{2}N^2 \frac{\Delta_2}{\omega_1 \omega_2}(\Delta_3-\Delta_1)	\ ,	\nn\\
  Q_3 - Q_2	&= \frac{1}{2}N^2 \frac{\Delta_1}{\omega_1 \omega_2}(\Delta_3-\Delta_2)	\ .
\label{eq: BH charges}
\end{align}
Using these relations, we find quite strikingly,
\begin{align}
  n^*_3 = - \frac{\Delta_1 \Delta_2}{\omega_1 \omega_2}N = \frac{2}{N}Q_3 + \frac{i}{\pi N}S\ .
\end{align}
In particular, $S$ is strictly positive (and grows like $N^2$), and thus $n_3^*$ lies strictly off the real line.
We thus find that a generic BPS black hole corresponds to an ensemble of giant graviton configurations of different wrapping numbers $n_3$.

\para 
Finally, to study this ensemble in more detail, we need to go back to the asymptotic expression
\begin{align}
  \frac{\I (\omega_i,\Delta_I)}{\I_\infty (\omega_i,\Delta_I)} \sim \int_0^\infty dn_3 \exp \left[-Nn_3 \Delta_3 - \frac{1}{2}n_3^2\frac{\omega_1 \omega_2 \Delta_3}{\Delta_1\Delta_2}\right]
\end{align}
but now fix the contour to remain on the real line, where it is summing over genuine giant graviton configurations. We can then consider the absolute value of the integrand as we integrate over $n_3$, for which we need the real parts of $\Delta_3$ and $\frac{\omega_1 \omega_2 \Delta_3}{\Delta_1\Delta_2}$. The most streamlined way to do this is to once again write everything in terms of microcanonical black hole quantities $S,J_i, Q_I$, which are manifestly real. This can be achieved using the relations
\begin{align}
  \frac{1}{2}N^2 \frac{\Delta_1 \Delta_2 \Delta_3}{\omega_1\omega_2} &= \left(\frac{1}{S-2\pi i Q_1}+\frac{1}{S-2\pi i Q_2} + \frac{1}{S-2\pi i Q_3} - \frac{1}{S+2\pi i J_1}- \frac{1}{S+2\pi i J_2}\right)^{-1}		\nn\\
  \frac{1}{2}N^2	\frac{\Delta_1 \Delta_2}{\omega_1\omega_2}	&= \frac{1}{2\pi i}\left(S-2\pi i Q_3\right)
\end{align}
Now, since the integrand has a rapidly oscillating phase as we pass over configurations, it is subtle to identify which giant graviton states dominate the black hole free energy. Nonetheless, it is a well-posed question to ask where in real $n_3$ space the integrand reaches its largest absolute value. This occurs at
\begin{align}
  n_3 = \hat{n}_3  = -\frac{2}{N} \text{Re}(\Delta_3) \left[\text{Re}\left(\frac{2\pi i \Delta_3}{S-2\pi i Q_3}\right)\right]^{-1}
\end{align}
which, written in terms of generic charges, takes a rather complicated form which we omit. As an example, in the case of a single-charge BPS black hole with $Q_1=Q_2=0$, we find
\begin{align}
  \hat{n}_3 = \frac{1}{N}\frac{J_1 J_2 \left(-4J_1^2 J_2^2+2N^2(J_1+ J_2)(J_1^2 + J_2^2)-N^4 (J_1 + J_2)^2\right)}{(J_1 + J_2)\left(2J_1^2 J_2^2 +N^2(J_1 + J_2)(J_1^2+J_2^2+3 J_1 J_2)\right)}
\end{align}
which does indeed scale like $N$, as $J_i \sim N^2$. Note that we have used the explicit form of the black hole entropy,
\begin{align}
  S=2\pi \sqrt{Q_1 Q_2+ Q_2 Q_3 + Q_3 Q_1 - \frac{1}{2}N^2 (J_1 + J_2)}
\end{align}
as well as eliminating $Q_3$ using the non-linear constraint which determines the fifth charge in terms of the four (\ref{eq: BH charges}), which in our conventions is
\begin{align}
  Q_1 Q_2 Q_3 + \frac{1}{2}N^2 J_1 J_2 = \left(Q_1 + Q_2 + Q_3 + \frac{1}{2}N^2\right)\left(Q_1 Q_2 + Q_2 Q_3 + Q_3 Q_1 - \frac{1}{2}N^2(J_1 +J_2)\right)
\end{align}
Further specialising to the case $J_1=J_2$, we have
\begin{align}
  \hat{n}_3 = \frac{2}{N} \frac{(Q_3+N^2)^2}{Q_3 - 2N^2}
\end{align}


\section{The AdS$_4\times \text{SE}_7$ case}
\label{sec: AdS4}

We now turn our attention to an analogous setup in AdS$_4$. 
By considering M2-branes sat at the tip of a Calabi-Yau 4-fold cone, one arrives at a duality between a three-dimensional $\N=2$ SCFT, and M-theory on AdS$_4\times \text{SE}_7$ for a given Sasaki-Einstein 7-fold SE$_7$. Then, the superconformal index of this theory admits an expansion in terms of contributions now from M5-brane giant gravitons wrapping cycles in the SE$_7$ in the AdS$_4\times \text{SE}_7$ dual geometry.

%

\subsection{Toric geometry and the master volume}

The geometrical aspects of SE$_7$, its metric deformation $Y_7$, and the associated master volume, have a qualitatively identical flavour to that of SE$_5$. Let us be fairly brief, and refer the reader to \cite{Gauntlett:2019roi} for the finer details.

\para
Now, the toric Sasaki-Einstein 7-manifold SE$_7$ is specified by a set of $D\ge 4$ vectors $\v_I\in \Z^4$, which can be taken of the form $\v_I = (v_I^1, v_I^2,v_I^3,1)$. The corresponding toric diagram is thus a compact, convex 3-dimensional polytope. A given such toric SE$_7$ is then a distinguished member of a broader family of geometries $Y_7$, specified by vectors $\v_I$ along with a Reeb vector $\b\in \R^4$ and $D$ K\"ahler moduli $\lambda_I$.

\para
Then, the \textit{master volume} is the volume of $Y_7$, and is given as follows. For each vertex $\v_I$ of the toric diagram, let $l_I$ denote the number of edges meeting there. Then, we let $\{\v_{I,1},\dots, \v_{I,l_I}\}$ denote the subset of $\{\v_1,\dots,\v_D\}$ of vertices that are connected by an edge to $\v_I$, ordered in an anti-clockwise manner when viewed from outside the toric diagram. Then, $\lambda_{I,k}$ is the K\"ahler modulus associated to the vertex $\v_{I,k}$, for each $k=1,\dots,l_I$. We finally define the scalar product of four vectors in $\R^4$, $(\u_1,\u_2,\u_3,\u_4) = \det(\u_1,\u_2,\u_3,\u_4)$. Then, the master volume is
\begin{align}
  \V(\b,\lambda_I) = - \frac{(2\pi)^4}{3!} \sum_{I=1}^D \lambda_I \sum_{k=2}^{l_I-1} \frac{X_{I,k}\tilde{X}_{I,k}}{(\v_I,\v_{I,k-1},\v_{I,k},\b)(\v_{I,l_I},\v_I,\v_{I,1},\b)(\v_{I,k},\v_{I,k+1},\v_I,\b)}\ ,
  \label{eq: AdS4 master volume}
\end{align}
where
\begin{align}
  X_{I,k}		= 	&- \lambda_I(\v_{I,k-1},\v_{I,k},\v_{I,k+1},\b)+ \lambda_{I,k-1}(\v_{I,k},\v_{I,k+1},\v_I,\b)		\nn\\
  					&- \lambda_{I,k}(\v_{I,k+1},\v_I,\v_{I,k-1},\b)+\lambda_{I,k+1}(\v_I,\v_{I,k-1},\v_{I,k},\b)	\ ,	\nn\\
  \tilde{X}_{I,k} = &-\lambda_I(\v_{I,1},\v_{I,k},\v_{I,l_I},\b) + \lambda_{I,1}(\v_{I,k},\v_{I,l_I},\v_I,\b)			\nn\\
  					&-\lambda_{I,k}(\v_{I,l_I},\v_I,\v_{I,1},\b) + \lambda_{I,l_I}(\v_I,\v_{I,1},\v_{I,k},\b)  \ .
\end{align}

\subsection{The giant graviton index}

We now pick some toric SE$_7$, which is specified by the set of fan vectors $\v_I$. One can then consider the three-dimensional $\N=2$ SCFT dual to AdS$_4\times \text{SE}_7$ \cite{Hanany:2008cd,Martelli:2008si}. This is the theory we now study.

\para
We can label states by their scaling dimension $\Delta$, their spin $J$, and their charges $Q_I$ under a $U(1)^D$ global symmetry\footnote{A particularly combination of these global rotations, determined by $F$-maximisation \cite{Jafferis:2010un}, forms the superconformal $U(1)_R$ R-symmetry.} . We choose a convention such that one of the two Poincar\'e supercharges $\Q$ carries charges
\begin{align}
  (+\tfrac{1}{2},-\tfrac{1}{2},+\tfrac{1}{2}) \qquad \text{under} \,\, (\Delta,J,Q_I)\ .
\end{align}
The superconformal index is then a trace over states on $S^2$, given by
\begin{align}
  \I(\omega,\Delta_I) =  \text{Tr} \left[e^{-\omega J - \Delta_I Q_I}\right]\ ,
  \label{eq: AdS4 index}
\end{align}
which is indeed an index for the BPS bound $\{\Q,\Q^\dagger\}\ge 0$ provided the chemical potentials satisfy
\begin{align}
  \omega - \sum_{I=1}^D \Delta_I = 2\pi i \quad  (\text{mod }4\pi i)\ .
  \label{eq: AdS4 linear relation}
\end{align}
Our interest is in the giant graviton expansion for $\I(\omega, \Delta_I)$. The relevant bulk objects are M5-branes wrapping the supersymmetric 5-cycles in SE$_7$ \cite{Arai:2020uwd}. There is such a cycle $S_I$ associated to each toric fan vector $\v_I$, so that the generic supersymmetric 5-cycle is specified by $D$ non-negative integers $\{n_I\}$, and wraps $S_I$ $n_I$ times. Thus, the giant graviton expansion expresses the superconformal index (\ref{eq: AdS4 index}) as a sum over contributions from M5-branes wrapping these 5-cycles,
\begin{align}
  \I(\omega,\Delta_I) = \I_\infty (\omega,\Delta_I) \sum_{n_I\in \Z_{\ge0}} e^{-N n_I \Delta_I} \,\I_{(n_1,\dots,n_D)} (\omega,\Delta_I)\ .
\label{eq: AdS4 GG expn}
\end{align}
This can be understood as a generalisation of the SE$_7 = S^7$ case discussed in \cite{Arai:2020uwd}.

\para
One is then interested in the large $n_I$ asymptotics of the giant graviton index $\I_{(n_1,\dots,n_D)}$. Schematically, $\I_{(n_1,\dots,n_D)}$ can be understood in much the same way as the D3-brane giant graviton index. It can be written as an integral over fugacities corresponding to a $U(n_1)\times\dots\times U(n_D)$ gauge symmetry. The integrand factorises into contributions from six-dimensional modes on each $S_I$, along with four-dimensional modes localised to any non-trivial double overlaps $S_I \cap S_J$ (corresponding to $\v_I$ and $\v_J$ being joined in the toric diagram), and two-dimensional modes localised to any non-trivial triple overlaps $S_I \cap S_J \cap S_K$ (corresponding to $\v_J$ being joined to both $\v_I$ and $\v_K$ in the toric diagram). One can then in principle seek a saddle-point approximation to this matrix model of the flavour of the `parallelogram ansatz' we saw in the D3-brane case.

\para
By a straightforward generalisation\footnote{We have fixed the pre-factor using the example of SE$_7 = S^7$, below.} of the result (\ref{eq: AdS5 result}), we conjecture that at large $n_I$,
\begin{align}
  \boxed{\log\I_{(n_1,\dots,n_D)}(\omega,\Delta_I) \sim -\frac{\omega^2}{4(2\pi)^4}\, \V(\b,\lambda_I)}
\label{eq: AdS4 result}
\end{align}
where we identify
\begin{align}
  \b= \beps = \sum_I \Delta_I \v_I,\qquad \lambda_I = n_I\ .
  \label{eq: AdS4 match}
\end{align}
In the remainder of this Section, let us verify this conjecture in the case of SE$_7 = S^7$, whereby the relevant asymptotics were already computed in \cite{Choi:2022ovw} by an anomaly-based approach. Now, $D=4$, with toric fan vectors $\v_1 = (0,0,0,1),\v_2 = (1,0,0,1),\v_3=(0,0,1,1),\v_4=(0,1,0,1)$, labelled such that $(\v_1,\v_2,\v_3,\v_4)=+1$.

\para
It was argued in \cite{Choi:2022ovw} that the leading asymptotic does indeed take the expected 6d/4d/2d form\footnote{Note that in \cite{Choi:2022ovw} there were certain sign ambiguities in identifying parameters of the six-/four-/two-dimensional contributions in terms of the chemical potentials $(\omega,\Delta_I)$. There is in fact no ambiguity, and this is just to do with determining the orientation of each five-/three-/one-dimensional submanifold of $S^7$ relative to that of $S^7$ as a whole. This is straightforward from the toric geometry perspective, although we omit details here.}
\begin{align}
  \log \I_{(n_1,n_2,n_3,n_4)}(\omega,\Delta_I) \sim \sum_{I=1}^4 \log\I^{(6d)}_{I}(\omega,\Delta_I) + \sum_{I<J} \log \I^{(4d)}_{I,J}(\omega,\Delta_I) + \sum_{I<J<K} \I^{(2d)}_{I,J,K}(\omega,\Delta_I)\ ,
\end{align}
where the tetrahedral toric diagram means that there are four double intersections $S_I \cap S_J \cong S^3$ and four triple intersections $S_I \cap S_J \cap S_K \cong S^1$. Adding up these contributions, one finds
\begin{align}
  \log \I_{(n_1,n_2,n_3,n_4)}(\omega,\Delta_I) \sim - \frac{\omega^2 \left(\sum_{I=1}^4 n_I \Delta_I\right)^3}{24 \Delta_1 \Delta_2 \Delta_3 \Delta_4}\ .
\label{eq: S7 asymptotic}
\end{align}
Let us then compare this against the right-hand-side of (\ref{eq: AdS4 result}). The master volume formula (\ref{eq: AdS4 master volume}) simplifies significantly in the case of SE$_7 = S^7$. We find
\begin{align}
  \V(\b,\lambda_I) = \frac{(2\pi)^4}{6} \frac{\left(-\lambda_1(\v_2,\v_3,\v_4,\b)+\lambda_2(\v_3,\v_4,\v_1,\b)- \lambda_3(\v_4,\v_1,\v_2,\b)+\lambda_4(\v_1,\v_2,\v_3,\b)\right)^3}{(\v_1,\v_2,\v_3,\b)(\v_2,\v_3,\v_4,\b)(\v_3,\v_4,\v_1,\b)(\v_4,\v_1,\v_2,\b)}\ .
\end{align}
Then, plugging in (\ref{eq: AdS4 match}), we have
\begin{align}
  \V(\b,\lambda_I) = \frac{(2\pi)^4}{6} \frac{\left(\sum_{I=1}^4 n_I \Delta_I\right)^3}{ \Delta_1 \Delta_2 \Delta_3 \Delta_4}\ .
\end{align}
We have thus verified (\ref{eq: AdS4 result}) in the case SE$_7 = S^7$.

\para
Let us finally make a comment on black holes. The large $N$ asymptotics of $\I(\omega,\Delta_I)$ have been studied in detail \cite{Choi:2019zpz,Choi:2019dfu,Nian:2019pxj,Bobev:2022wem}. In a suitable regime of chemical potential space, one finds
\begin{align}
  \log\I(\omega,\Delta_I) \sim \pm\frac{4 \sqrt{2}\,i}{3} N^{3/2} \frac{\sqrt{\Delta_1\Delta_2\Delta_3\Delta_4}}{\omega}\ ,
\end{align}
with a sign dictated by the chemical potential regime. This expression coincides with minus the on-shell action of the dual BPS rotating black hole. Alternatively, we can interpret the right-hand-side as the entropy function of this black hole \cite{Choi:2018fdc}. Then, it was shown in \cite{Choi:2022ovw} using the asymptotic (\ref{eq: S7 asymptotic}) that one recovers precisely this asymptotic from a saddle-point approximation of the sum over the $\{n_I\}$ in (\ref{eq: AdS4 GG expn}) (after approximating this sum by an integral over continuous $n_I$).



\section{Discussion and open questions}
\label{sec: discussion}
In this paper, we uncovered a new aspect of the holographic duality enjoyed by the backgrounds AdS$_5\times \text{SE}_5$ and AdS$_4\times \text{SE}_7$. We demonstrated a correspondence between on one hand the asymptotics of the giant graviton contributions to the superconformal index of the dual SCFT, and on the other, the master volume of a particular family of metric deformations of SE$_5$ and SE$_7$, respectively. In the large-$N$ limit, the superconformal index can then be recast as an integral over the K\"{a}hler parameters of this geometry, and the leading saddle-point contribution is obtained by stationarising the master volume function in the presence of appropriate sources. The asymptotic index degeneracy at fixed charge is obtained by further stationarising with respect to the components of the Reeb vector (along with $\omega_1,\omega_2$) and reproduces the known match to the entropy of rotating black holes in the AdS dual. At a purely mathematical level, we have thus found a geometrical reformulation of a computation in the boundary theory and it is natural to seek a physical interpretation for this geometry in terms of bulk gravity. 
Indeed, in the related context of topologically-twisted indices, precisely such an interpretation exists. In this context, the same metrics appear as a factor in the near-horizon geometry of static, magnetically charged black strings and black holes (in the $D=4$ \cite{Gauntlett:2018dpc}  and $D=3$ \cite{Gauntlett:2019roi} cases respectively). 
Our results suggest an extension of these ideas to the near-horizon geometry of $\frac{1}{16}$-BPS rotating black holes. However, as we now discuss, there are also novel features of our set-up.
\para 
The master volume formalism of GMS considers the same family of metrics on  
$Y_{5}\simeq \text{SE}_5$ and $Y_{7}\simeq SE_{7}$. The near-horizon geometry of magnetic black strings (in AdS$_{5}$) and black holes (in AdS$_{4}$) is obtained by fibering these spaces over $\mathbb{CP}^{1}$. More precisely, the near-horizon metric on the resulting geometry is obtained by stationarizing an action\footnote{It is stated in \cite{Gauntlett:2019roi}, as well as related work, that the K\"ahler parameters $\lambda_I$ are fixed by imposing flux constraints that schematically take the form $M_I - \sum_J \partial^2 \V(\b,\lambda)/\partial \lambda_I \partial \lambda_J=0$, for some fluxes $M_I$. Our observation is a simple one, that these constraints can equivalently be stated as the Euler-Lagrange equations of the action $S=M_I \lambda_I - \sum_J \partial \V(\b,\lambda)/\partial \lambda_J$. In particular, this action depends on the first derivative of the master volume $\V$, rather than on $\V$ itself as we find here in the rotating case.} over both the K\"{a}hler parameters $\{\lambda_{I}\}$ and the components of the Reeb vector $\mathbf{b}$. Although different metrics in the variational family can be probed by introducing background fluxes and charges, only the stationary points of the action correspond to supergravity solutions. In contrast, our results suggest a physical interpretation for the metric for all values of the parameters: it should correspond to the near-horizon geometry of a particular configuration of giant gravitons. Of course, our results hold in the context of the superconformal index which has a known expansion in terms of giant gravitons. Further progress in this direction depends on establishing a corresponding giant graviton index for the topologically-twisted index\footnote{A particularly interesting possibility is that this connection holds at the level of the individual holomorphic blocks which can be glued to form either superconformal or topologically twisted indices and the parallel decomposition into gravitational blocks on bulk side. A giant graviton expansion for holomorphic blocks is currently under investigation.}.     
\para 
As in the static magnetically-charged cases considered by GMS, the entropy of $\frac{1}{16}$-BPS rotating black holes in AdS$_{5}\times \text{SE}_5$ (and AdS$_{4}\times \text{SE}_{7}$) is obtained by stationarising over K\"{a}hler moduli and the components of the Reeb vector. In the rotating case, the resulting saddle points occur at complex values of the parameters and the resulting geometry should be interpreted as a complex saddle-point of the Euclidean gravitational path integral. It would be interesting to identify the resulting complex geometry explicitly. 

\para   
Another interesting aspect of the correspondence described above relates to the high degree of redundancy in the description of giant graviton states which leads to large cancellations between different sectors\footnote{As emphasised recently in \cite{Lee:2022vig,Lee:2023iil}, this reflects an underlying gauge invariance in the bulk description closely related to the existence of trace constraints in the boundary theory.}. One manifestation of this is the existence of many different versions of the giant graviton expansion, including some in which the number of summation variables can be reduced from $D$ to $D-2$. This reduction procedure was used in \cite{Fujiwara:2023bdc} to simplify the derivation of large-$N$ asymptotics. As we have seen above, exactly the same redundancy appears on our reformulation of the problem due to a well known gauge invariance of the master volume function. It would be interesting to understand the appearance of this gauge symmetry directly from a bulk point of view.                  
\para
Finally, there have been other attempts to relate the terms in the giant graviton expansion (\ref{eq: GG expn intro}) to features of classical geometries. Notably, \cite{Deddo:2024liu} takes a more enumerative approach, suggesting that $\I_{(n_1,\dots,n_D)}$ can be recovered from counting bubbling geometries. It would be very interesting to understand if and how their perspective and ours are related.

\subsection*{Acknowledgements}

We are grateful to Alejandra Castro, Jerome Gauntlett, Dario Martelli, and Sameer Murthy for helpful comments on a draft of this paper. We would like to also thank Yosuke Imamura and Seok Kim for valuable discussions. 

\para 
The work of H.Y.C. was supported in part by Ministry of Science and Technology (MOST) through the grant 110-2112-M-002-006-. H. Y. C. and S. M. would also like to thank Yukawa Institute for Theoretical Physics, Kyoto University for warm hospitality during the completion of this work.
The work of N. D. and R. M. was supported by STFC consolidated grant ST/T000694/1. The work of N. D. was also supported by a TSVP Visiting Fellowship at the Okinawa Institute of Science and Technology. 
The work of S. M. was supported by JSPS Grant-in-Aid for Scientific Research (C) \#22K03598.
The work of R. M. was also supported by David Tong's Simons Investigator Award, and the UK Engineering and Physical Sciences Research council grant EP/Z000106/1.
The work of C. \c{S}. was supported by the European Union's Horizon Europe programme under grant agreement No. 101109743, project Quivers. C. \c{S}. is grateful to Okinawa Institute of Science and Technology (OIST) Theoretical Sciences Visiting Program (TSVP) and the DAMTP, University of Cambridge for their warm hospitality during part of this research.

\newpage 
\appendix 
\part*{Appendices}

\section{Asymptotics of the D3-brane giant graviton index}
\label{app: asymptotics}

In this Appendix, we provide details of the derivation of the main results of Section \ref{subsec: AdS5 GGI}.

\subsection{Four-dimensional contributions}

We first study the four-dimensional contributions $F_I^{(4d)}$ appearing in (\ref{eq: GG index schematic}). This captures the modes living on $n_I$ D3-branes wrapped on the supersymmetric 3-cycle $S_I\subset \text{SE}_5$. Perhaps the most straightforward way to derive an expression for $F_I^{(4d)}$ is by performing a particular orbifold projection of (the integrand of) the $S^1\times S^3$ superconformal index of $U(n_I)$ $\N=4$ super-Yang-Mills \cite{Arai:2019aou,Fujiwara:2023bdc,Kim:2024ucf,Benini:2011nc,Razamat:2013opa}. This index is graded with respect to angular momenta $J_1, J_2$ and R-charges $Q_1, Q_2, Q_3$, where our conventions are such that the sixteen Poincar\'e supercharges carry charges $\pm \frac{1}{2}$ under each of these generators with the product of all five signs being $+$. We consider the index defined with respect to the supercharge with charges $(-\frac{1}{2},-\frac{1}{2},+\frac{1}{2},+\frac{1}{2},+\frac{1}{2})$ under $(J_1,J_2,Q_1,Q_2,Q_3)$. The index then depends on angular momentum chemical potentials $\omega_{1,2}^{(I)}$ and R-symmetry chemical potentials $\Delta_{1,2,3}^{(I)}$, subject to the constraint
\begin{align}
  \omega_1^{(I)} + \omega_2^{(I)} - \Delta_1^{(I)}- \Delta_2^{(I)}- \Delta_3^{(I)} = 2\pi i \quad (\text{mod }4\pi i)\ .
\label{eq: SUSY constraint appendix}
\end{align}
The values of these chemical potentials are fixed by the chemical potentials $\omega_{1,2}$ and $\Delta_I$ of the $\N=1$ index $\I(\omega_i, \Delta_I)$. The precise map is fixed by the toric data, and is given in (\ref{eq: 4d chem pots}).

\para 
The toric data then fixes the orbifold we need to take. It is generated by
\begin{align}
  \exp\left(\frac{2\pi i}{\v_{I-1} \cdot (\v_{I}\times \v_{I+1})}\left( J_1 + \alpha_I J_2 + (1 + \alpha_I)Q_3\right)\right)\ .
  \label{eq: orbifold generator}
\end{align}
Importantly, this generator commutes with the supercharge with respect to which we have defined out superconformal index. 
Here, $0< \alpha_I < k_I$ is a specific integer that is coprime to  $k_I= \v_{I-1} \cdot (\v_{I}\times \v_{I+1})$. It follows that this defines a free $\Z_{k_I}$ orbifold. There is an algorithm to determine $\alpha_I$. Roughly speaking, one first forms a parallelogram in $\R^2$ whose sides are $\left(\v_{I-1}-\v_I\right)$ and $\left(\v_I - \v_{I+1}\right)$, and whose area is thus $k_I$. Then, there are precisely $k_I-1$ lattice points lying within this parallelogram, and each can be associated with a unique non-identity element of $\Z_{k_I}$. In particular, there always exists one such point which corresponds to the generating element (\ref{eq: orbifold generator}). The coordinates of this point relative to the parallelogram determines $\alpha_I$. More details of this algorithm can be found in \cite{Arai:2019aou}. It will turn out that for our purposes, we will not need an explicit form for $\alpha_I$. 

\para
One needs to also think about twisted sectors. These correspond precisely to the $k_I^{n_I}$ sectors of different $U(n_I)$ gauge holonomy around the non-contractible cycle in $S_I \cong S^3/\Z_{k_I}$. An equivalent way to think about these twisted sectors is to note that we can always absorb a flat connection with non-trivial holonomy into fields, at the expense of introducing a gauge rotation to the orbifold generator (\ref{eq: orbifold generator}). So, different twisted sectors correspond to different ways in which our orbifold action can act within the gauge group. It was argued in \cite{Fujiwara:2023bdc} that sectors of different holonomy could be related to sectors of differing baryonic charges in the $\N=1$ index $\I(\omega_i,\Delta_I)$. Our interest is in the $n_I\to \infty$ asymptotics. We will soon see that at leading order in this regime, every term in this sum over holonomies contributes equally to $F_I^{(4d)}$, and so the only impact of having to sum over these holonomies is the addition of a $n_I \log k_I$ correction to the large $n_I$ asymptotics of $\log \I_{(n_1,\dots,n_D)}$, which will turn out to be subleading, since $\log \I_{(n_1,\dots,n_D)}\sim n_I^2$.  

\para
Let us focus on some fixed $I\in \{1,\dots,D\}$. It is useful to define fugacities $p_{1,2}$ and $ x_{1,2,3}$, and rescaled chemical potentials $\sigma_{1,2}$ and $\nu_{1,2,3}$, by
\begin{align}
  p_{1,2} = e^{2\pi i \sigma_{1,2}} = e^{-\omega^{(I)}_{1,2}},\qquad x_{1,2,3} = e^{2\pi i \nu_{1,2,3}} = e^{- \Delta^{(I)}_{1,2,3}}\ .
  \label{eq: appendix new chems}
\end{align}
The constraint (\ref{eq: SUSY constraint appendix}) is recast as
\begin{align}
  \sigma_1 + \sigma_2 - \nu_1 - \nu_2 - \nu_3 = 1 \quad (\text{mod }2)\ .
\end{align}
For the sake of presentation, it will be useful to fix without loss of generality the right-hand-side of (\ref{eq: linear relation}) to equal $-2\pi i$, which using (\ref{eq: 4d chem pots}) implies that the right-hand-side of (\ref{eq: SUSY constraint appendix}) equals $+2\pi i$, and thus
\begin{align}
  \sigma_1 + \sigma_2 - \nu_1 - \nu_2 - \nu_3 = -1\ .
\label{eq: app local lin rel}
\end{align}
We also define gauge fugacities $z_{I,a}$ by
\begin{align}
  z_{I,a} = e^{2\pi i u_{I,a}}\ .
\end{align}
For ease of notation, let us temporarily drop the $I$ subscripts on $F_I^{(4d)},n_I,k_I,\alpha_I, u_I,z_{I,a}$, always in the knowledge that we are working at the $I^\text{th}$ node.
Then, we have
\begin{align}
  F^{(4d)} = \sum_{h=(h_1,\dots,h_n)} F^{(4d)}_h\ ,
\label{eq: holonomy sum}
\end{align}
where the sum is taken over holonomies $(e^{2\pi i h_1/k},\dots,e^{2\pi i h_n/k})\in (\Z_k)^n$, where $0\le h_a < k$.  We have then 
\begin{align}
  F^{(4d)}_h = \mu_h(z)\,\text{Pexp}\left[\frac{1}{k}\sum_{l=0}^{k-1} \left(1-\frac{(1-x_1)(1-x_2)(1-\omega^{(1+\alpha)l}x_3)}{(1-\omega^{ l}p_1)(1-\omega^{\alpha l}p_2)}\right) \sum_{a,b=1}^n \omega^{(h_a-h_b)l} \frac{z_a}{z_b} \right] \ ,
\end{align}
in terms of $\omega=e^{2\pi i/k}$. The prefactor $\mu_h(z)$ is the Haar measure for the subgroup of $U(n_I)$ that is preserved by the holonomy $h$,
\begin{align}
  \mu_h(z) = \prod_{\substack{(a,b)\in K_h \\a\neq b}} \left(1-\frac{z_a}{z_b}\right)\ ,
\end{align}
where here $(a,b)\in K_h\subset \{1,\dots,n\}^2$ precisely if $h_a=h_b$. For example, if $n=3$ and $h_1=h_2\neq h_3$, we have $K_h=\{(1,1),(2,2),(3,3),(1,2),(2,1)\}$.

\para
Let us then manipulate the plethystic exponential. For the first term, we have
\begin{align}
  \text{Pexp}\left[\frac{1}{k}\sum_{l=0}^{k-1} \sum_{a,b=1}^n \omega^{(h_a-h_b)l} \frac{z_a}{z_b} \right] &= \text{Pexp}\left[ \sum_{(a,b)\in K_h}  \frac{z_a}{z_b} \right]			\nn\\
  &= \mu_h(z)^{-1}\ ,
\end{align}
where in reaching the final expression, we note that the diagonal term is automatically removed by virtue of the definition of the plethystic exponential,
\begin{align}
  f(x) = \sum_{n=0}^\infty a_n x^n \quad \implies \quad \text{Pexp}[f(x)] = \exp \left(\sum_{n=1}^\infty \frac{f(x^n)-f(0)}{n}\right) = \prod_{n=1}^\infty \left(1-x^n\right)^{-a_n}\ .
\end{align}
So we're left with the second term. We play a few tricks here.
\begin{align}
  &\text{Pexp}\left[-\frac{1}{k}\sum_{l=0}^{k-1} \left(\frac{(1-x_1)(1-x_2)(1-\omega^{(1+\alpha)l}x_3)}{(1-\omega^{l}p_1)(1-\omega^{\alpha l}p_2)}\right) \sum_{a,b=1}^n \omega^{(h_a-h_b)l} \frac{z_a}{z_b} \right] 		\nn\\
  & = \text{Pexp}\left[-\frac{1}{k}\sum_{a,b=1}^n\sum_{l,m_1,m_2=0}^{k-1} \left(\frac{(1-x_1)(1-x_2)(1-\omega^{(1+\alpha)l}x_3)}{(1-p_1^k)(1-p_2^k)}\right) p_1^{m_1} p_2^{m_2}  \omega^{(h_a-h_b+m_1+\alpha m_2)l} \frac{z_a}{z_b} \right]		\nn\\
  &= \text{Pexp}\left[-\frac{(1-x_1)(1-x_2)}{(1-p_1^k)(1-p_2^k)}\right.	\nn\\
  &\hspace{20mm}\left. \times\sum_{a,b=1}^n \sum_{m_1,m_2=0}^{k-1} \left(\delta_{h_a-h_b+m_1 + \alpha m_2,0}- \delta_{h_a-h_b+(m_1+1) + \alpha (m_2+1),0}x_3\right) p_1^{m_1}p_2^{m_2}\frac{z_a}{z_b} \right]		\nn\\
  &= \text{Pexp}\left[-\frac{(1-x_1)(1-x_2)}{(1-p_1^k)(1-p_2^k)}\right.	\nn\\
  &\hspace{20mm}\left. \times\sum_{a,b=1}^n \sum_{m_1,m_2=0}^{k-1} \delta_{h_a-h_b+m_1 + \alpha m_2,0} \left(p_1^{m_1}p_2^{m_2}\frac{z_a}{z_b} - x_3 p_1^{k-1-m_1}p_2^{k-1-m_2}\frac{z_b}{z_a} \right)\right]	\ ,
\label{eq: plethystic manipulations}
\end{align}
where here, $\delta_{n,0}=1$ if $n=0$ mod $k$, and is zero otherwise. Putting everything together, we find
\begin{align}
  &F_h^{(4d)}=  \prod_{a\neq b} \prod_{m=0}^{k-1}\nn\\
  &  \frac{\Gamma\!\left(u_{ab}+ \llbracket h_b-h_a-\alpha m \rrbracket\sigma_1 + m \sigma_2+\nu_1,k\sigma_1,k\sigma_2\right)\Gamma\!\left(u_{ab}+\llbracket h_b-h_a-\alpha m \rrbracket\sigma_1 + m \sigma_2+\nu_2,k\sigma_1,k\sigma_2\right)}{\Gamma\!\left(u_{ab}+\llbracket h_b-h_a-\alpha m \rrbracket\sigma_1 + m \sigma_2+\nu_1+\nu_2,k\sigma_1,k\sigma_2\right)\Gamma\!\left(u_{ab}+\llbracket h_b-h_a-\alpha m \rrbracket\sigma_1 + m \sigma_2,k\sigma_1,k\sigma_2\right)}\ ,
\label{eq: 4d piece as gammas}
\end{align}
where $u_{ab}=u_a-u_b$, and we have used the elliptic gamma function\footnote{Note that the infinite product representation here is only valid when $\sigma,\tau$ lie in the upper-half-plane. One can nonetheless define infinite product expressions for $\Gamma(x,\sigma,\tau)$ for all $\sigma,\tau$ away from the real axis, by imposing the relations $\Gamma(x,-\sigma,\tau) = 1/\Gamma(x+\sigma,\sigma,\tau)$ and $\Gamma(x,\sigma,-\tau)=1/\Gamma(x+\tau,\sigma,\tau)$ \cite{felder2000elliptic}. In contrast, the Plethystic expression we present here---which after all is what actually appears in (\ref{eq: plethystic manipulations})---makes sense for all $\sigma,\tau$ away from the real line, and reproduces the correct infinite product in the four sectors $\text{Im}\, \sigma,\text{Im}\,\tau \gtrless 0$. }
\begin{align}
  \Gamma(x,\sigma,\tau)= \text{Pexp}\left[\frac{e^{2\pi i x} - e^{2\pi i\left(\sigma + \tau - x\right)}}{(1-e^{2\pi i \sigma})(1-e^{2\pi i \tau})}\right] = \prod_{n,m\ge 0}\frac{1-e^{2\pi i\left(-x+(n+1)\sigma + (m+1)\tau\right)}}{1-e^{2\pi i\left(x+n\sigma + m\tau\right)}} \ .
\label{eq: Gamma def}
\end{align}
Furthermore, for any integer $n$ we define $\llbracket n \rrbracket$ to be the unique integer in $\{0,\dots,k-1\}$ such that $\llbracket n \rrbracket = n$ (mod $k$). In particular, $\llbracket -n \rrbracket = k - \llbracket n \rrbracket$ unless $n=0$.


%

\para
Note that for a special choice of holonomy, in which the gauge group is broken $U(n_I)\to U(n_I/k)^k$, our result (\ref{eq: 4d piece as gammas}) for generic holonomies degenerates to a previous result of \cite{Kim:2024ucf}.

\para
Next, we leverage some properties of the elliptic gamma function. Amongst the $SL(3,\Z)$ modular transformations of $\Gamma(x,\sigma,\tau)$, we will use
\begin{align}
  \Gamma(x,\sigma,\tau) &= e^{-i\pi P_+(x,\sigma,\tau)}\, \Gamma\!\left(-\frac{x+1}{\sigma},-\frac{1}{\sigma},-\frac{\tau}{\sigma}\right)\Gamma \!\left(\frac{x}{\tau}, - \frac{1}{\tau}, \frac{\sigma}{\tau}\right)		\nn\\
  &= e^{-i \pi P_-(x,\sigma,\tau)}\, \Gamma\!\left(-\frac{x}{\sigma},-\frac{1}{\sigma},-\frac{\tau}{\sigma}\right)\Gamma \!\left(\frac{x-1}{\tau}, - \frac{1}{\tau}, \frac{\sigma}{\tau}\right)	\ ,
\label{eq: mod transform of gamma}
\end{align}
where
\begin{align}
  P_\pm (x,\sigma,\tau) &= \frac{1}{3\sigma\tau}x^3 - \frac{\sigma + \tau \mp 1}{2\sigma \tau} x^2 + \frac{\sigma^2 + \tau^2 + 3\sigma \tau \mp 3\sigma \mp 3\tau + 1}{6 \sigma \tau}x \nn\\
  &\qquad \pm \frac{1}{12}(\sigma + \tau \mp 1)\left(\frac{1}{\sigma}+\frac{1}{\tau} \mp 1\right)\ .
\end{align}
%
One can use these transformations to show that
\begin{align}
  &\hspace{-10mm}\frac{\Gamma(x+\gamma+\alpha,\sigma,\tau)\Gamma(x+\gamma+\beta,\sigma,\tau)\Gamma(-x+\gamma+\alpha,\sigma,\tau)\Gamma(-x+\gamma+\beta,\sigma,\tau)}{\Gamma(x+\gamma+\alpha+\beta,\sigma,\tau)\Gamma(x+\gamma,\sigma,\tau)\Gamma(-x+\gamma+\alpha+\beta,\sigma,\tau)\Gamma(-x+\gamma,\sigma,\tau)} 		\nn\\
  &\hspace{-10mm} = \exp\left(\frac{2\pi i \alpha \beta (\alpha+\beta-\sigma - \tau + 2\gamma - 1)}{\sigma \tau}\right)				\nn\\
  &\hspace{-10mm}\times \left(\frac{
  \Gamma\!\left(-\frac{x+\gamma+\alpha}{\sigma},-\frac{1}{\sigma},-\frac{\tau}{\sigma}\right)
  \Gamma\!\left(-\frac{x+\gamma+\beta}{\sigma},-\frac{1}{\sigma},-\frac{\tau}{\sigma}\right)
  \Gamma\!\left(-\frac{-x+\gamma+\alpha}{\sigma},-\frac{1}{\sigma},-\frac{\tau}{\sigma}\right)
  \Gamma\!\left(-\frac{-x+\gamma+\beta}{\sigma},-\frac{1}{\sigma},-\frac{\tau}{\sigma}\right)
  }{
  \Gamma\!\left(-\frac{x+\gamma+\alpha+\beta}{\sigma},-\frac{1}{\sigma},-\frac{\tau}{\sigma}\right)
  \Gamma\!\left(-\frac{x+\gamma}{\sigma},-\frac{1}{\sigma},-\frac{\tau}{\sigma}\right)
  \Gamma\!\left(-\frac{-x+\gamma+\alpha+\beta}{\sigma},-\frac{1}{\sigma},-\frac{\tau}{\sigma}\right)
  \Gamma\!\left(-\frac{-x+\gamma}{\sigma},-\frac{1}{\sigma},-\frac{\tau}{\sigma}\right)
  }  \right)		\nn\\
  &\hspace{-10mm}\times \left(\frac{
  \Gamma\!\left(\frac{x+\gamma+\alpha-1}{\tau},-\frac{1}{\tau},\frac{\sigma}{\tau}\right)
  \Gamma\!\left(\frac{x+\gamma+\beta-1}{\tau},-\frac{1}{\tau},\frac{\sigma}{\tau}\right)
  \Gamma\!\left(\frac{-x+\gamma+\alpha-1}{\tau},-\frac{1}{\tau},\frac{\sigma}{\tau}\right)
  \Gamma\!\left(\frac{-x+\gamma+\beta-1}{\tau},-\frac{1}{\tau},\frac{\sigma}{\tau}\right)
  }{
  \Gamma\!\left(\frac{x+\gamma+\alpha+\beta-1}{\tau},-\frac{1}{\tau},\frac{\sigma}{\tau}\right)
  \Gamma\!\left(\frac{x+\gamma-1}{\tau},-\frac{1}{\tau},\frac{\sigma}{\tau}\right)
  \Gamma\!\left(\frac{-x+\gamma+\alpha+\beta-1}{\tau},-\frac{1}{\tau},\frac{\sigma}{\tau}\right)
  \Gamma\!\left(\frac{-x+\gamma-1}{\tau},-\frac{1}{\tau},\frac{\sigma}{\tau}\right)
  }  \right)
\end{align}
%
We apply this identity to groups of elliptic gamma functions appearing in the product (\ref{eq: 4d piece as gammas}). In particular, we group the $(a,b,m)$ factor with the $(b,a,\llbracket -m\rrbracket)$ factor. We find
\begin{align}
  F_h^{(4d)} = \Upsilon^{(4d)}_h \tilde{F}_h^{(4d)}\ .
\label{eq: Fh transformed}
\end{align}
We leave the pieces $\tilde{F}_h^{(4d)}$ until later. For now, our focus is on the modular kernel $\Upsilon^{(4d)}_h$. We find
\begin{align}
  \Upsilon^{(4d)}_h &= \exp\left(\frac{2\pi i \nu_1 \nu_2}{k^2 \sigma_1 \sigma_2} \sum_{a<b}\left(-k\sigma_1 - k \sigma_2+\sum_{m=0}^{k-1} \left(\nu_1 +\nu_2-1\right)\right)  \right)		\nn\\
  &= \exp \left(\frac{\pi i}{k}n(n-1)\frac{\nu_1 \nu_2}{\sigma_1 \sigma_2}\left(\nu_1 + \nu_2 - \sigma_1 - \sigma_2 - 1\right)\right)		\nn\\
  &= \exp\left(-\frac{\pi i}{k}n(n-1) \frac{\nu_1 \nu_2 \nu_3}{\sigma_1 \sigma_2}\right)	\ .	
  \label{eq: Ups h}
\end{align}
In reaching the final equality, we have used (\ref{eq: app local lin rel}).

\para
Now we perform the sum over holonomies (\ref{eq: holonomy sum}). The key point is that every term in this sum exhibits the \textit{same} modular kernel, and so we find
\begin{align}
  F_I^{(4d)} = \Upsilon^{(4d)}_I \tilde{F}_I^{(4d)}\ ,
\end{align}
where
\begin{align}
  \tilde{F}_I^{(4d)} = \sum_h \tilde{F}_{I,h}^{(4d)}\ ,
  \label{eq: holonomy decomp}
\end{align}
with the terms in this sum being the transformed functions appearing in (\ref{eq: Fh transformed}). Rewriting (\ref{eq: Ups h}) in terms of our unscaled chemical potentials at the $I^\text{th}$ node, we have
\begin{align}
  \Upsilon_I^{(4d)} = \exp\left(\frac{n_I(n_I-1)}{2k_I} \frac{\Delta_1^{(I)}\Delta_2^{(I)}\Delta_3^{(I)}}{\omega_1^{(I)}\omega_2^{(I)}}\right)\ ,
\label{eq: 4d modular kernel explicitly}
\end{align}
and thus at large $n_I$,
\begin{align}
  \log \Upsilon_I^{(4d)} \sim \frac{n_I^2}{2k_I} \frac{\Delta_1^{(I)}\Delta_2^{(I)}\Delta_3^{(I)}}{\omega_1^{(I)}\omega_2^{(I)}}\ ,
\end{align}
and we establish the result (\ref{eq: 4d free energy}) stated in the main text.

\subsection{Two-dimensional contributions}

We next study the $F_{I,I+1}^{(2d)}$ appearing in (\ref{eq: GG index schematic}), following \cite{Fujiwara:2023bdc,Kim:2024ucf}. We have
\begin{align}
  F_{I,I+1}^{(2d)} = \prod_{a=1}^{n_I} \prod_{b=1}^{n_{I+1}} &\left[ \frac{\vartheta\! \left(u_{I,a} - u_{I+1,b}-\frac{1}{2}-\frac{1}{4\pi i}(\beps\cdot(\v_I\times \v_{I+1})+\omega_1-\omega_2),-\frac{\beps\cdot(\v_I\times \v_{I+1})}{2\pi i}\right) }{\vartheta\! \left(u_{I,a} - u_{I+1,b}-\frac{1}{2}-\frac{1}{4\pi i}(\beps\cdot(\v_I\times \v_{I+1})-\omega_1-\omega_2),-\frac{\beps\cdot(\v_I\times \v_{I+1})}{2\pi i}\right)} \right.			\nn\\
  & \quad \left.\times \frac{\vartheta\! \left(-u_{I,a} + u_{I+1,b}-\frac{1}{2}-\frac{1}{4\pi i}(\beps\cdot(\v_I\times \v_{I+1})+\omega_1-\omega_2),-\frac{\beps\cdot(\v_I\times \v_{I+1})}{2\pi i}\right) }{\vartheta\! \left(-u_a^I + u_b^{I+1}-\frac{1}{2}-\frac{1}{4\pi i}(\beps\cdot(\v_I\times \v_{I+1})-\omega_1-\omega_2),-\frac{\beps\cdot(\v_I\times \v_{I+1})}{2\pi i}\right)} \right]\ ,
\label{eq: 2d contribution in full}
\end{align}
written in terms of the $q$-theta function,
\begin{align}
  \vartheta(x,\tau) = \prod_{n=0}^\infty  \left(1-e^{2\pi i(-x+(n+1)\tau)} \right)\left(1-e^{2\pi i(x+n\tau)} \right)\ .
\end{align}
 We once again play some modular games. We make use of the identity
\begin{align}
  \vartheta(x,\tau) = e^{-i\pi B(x,\tau)}\vartheta\!\left(\frac{x}{\tau},-\frac{1}{\tau}\right)\ ,
\label{eq: mod transform theta function}
\end{align}
where
\begin{align}
  B(x,\tau) = \frac{1}{\tau}x^2 + \left(\frac{1}{\tau}-1\right)x + \frac{1}{6}\left(\tau+\frac{1}{\tau}\right) - \frac{1}{2}\ .
\end{align}
%
From this, we find the identity
\begin{align}
  &\frac{\vartheta(x +\alpha+\gamma,\beta )\vartheta(-x +\alpha+\gamma,\beta )}{\vartheta(x -\alpha+\gamma,\beta )\vartheta(-x -\alpha+\gamma,\beta )}		= \exp\left(\frac{4\pi i \alpha(\beta-2\gamma-1)}{\beta}\right) \frac{\vartheta(\frac{x +\alpha+\gamma}{\beta},-\frac{1}{\beta} )\vartheta(\frac{-x +\alpha+\gamma}{\beta},-\frac{1}{\beta} )}{\vartheta(\frac{x -\alpha+\gamma}{\beta},-\frac{1}{\beta} )\vartheta(\frac{-x -\alpha+\gamma}{\beta},-\frac{1}{\beta} )}		\ .
\end{align}
Applying this identity to each term in the product (\ref{eq: 2d contribution in full}), we have
\begin{align}
  F_{I,I+1}^{(2d)} = \modker_{I,I+1}^{(2d)} \tilde{F}_{I,I+1}^{(2d)}\ .
\end{align}
We once again leave analysis of the transformed contribution $\tilde{F}_{I,I+1}^{(2d)}$ until later, and focus on the modular kernel $\modker_{I,I+1}^{(2d)}$. We find
\begin{align}
  \modker_{I,I+1}^{(2d)} = \exp\left(-n_I n_{I+1} \frac{\omega_1 \omega_2}{\beps\cdot(\v_I\times \v_{I+1})}\right)\ ,
\end{align}
thus establishing the result (\ref{eq: 2d result main text}). 

\subsection{The parallelogram ansatz}

With the nice bits extracted---the modular kernels $\modker_I^{(4d)}$ and $\modker_{I,I+1}^{(2d)}$---we're left with a beast. Namely, the integral (\ref{eq: remaining integral}). Using the holonomy decomposition (\ref{eq: holonomy decomp}) for each four-dimensional factor, we have
\begin{align}
  \tilde{\I}_{(n_1,\dots,n_D)}(\omega_i, \Delta_I) = \sum_{h^{(1)},\dots,h^{(D)}} \tilde{\I}_{(n_1,\dots,n_D)}(\omega_i, \Delta_I; h^{(I)})\ .
\end{align}
Here, $h^{(I)}\in (\Z_{k_I})^{n_I}$ specifies the gauge holonomy of the $I^\text{th}$ four-dimensional contribution. Thus, this sum has $k_1^{n_1}\dots k_D^{n_D}$ terms. We will shortly argue that in a suitable regime of chemical potentials $(\omega_i, \Delta_I)$, as we take $n_I\to \infty$ we have 
\begin{align}
  \tilde{\I}_{(n_1,\dots,n_D)}(\omega_i, \Delta_I;h^{(I)}) \sim 1\ ,
  \label{eq: single holo order 1}
\end{align}
and thus 
\begin{align}
  \log \tilde{\I}_{(n_1,\dots,n_D)}(\omega_i, \Delta_I) \sim \sum_I n_I \log k_I\ ,
\end{align}
as stated in the main text.

\para
So let us demonstrate (\ref{eq: single holo order 1}). Explicitly, we have
\begin{align}
  &\tilde{\I}_{(n_1,\dots,n_D)}(\omega_i, \Delta_I; h^{(I)}) \nn\\
  &\qquad = \int \left(\prod_{I=1}^D d\,^{n_I}u_I \right) \left(\prod_{I=1}^D \tilde{F}_{I,h^{(I)}}^{(4d)}(\omega_i, \Delta_I; u_I)\right)\left(\prod_{I=1}^D \tilde{F}_{I,I+1}^{(2d)}(\omega_i, \Delta_I; u_I, u_{I+1})\right)\ .
\label{eq: integral for analysis}
\end{align}
To write this integrand out explicitly, it is helpful to define the variables
\begin{align}
  \rho_{I,I+1}:= - \frac{1}{2\pi i} \beps \cdot(\v_I\times \v_{I+1})\ .
\end{align}
Note that $(\omega_1+\omega_2)$ is related to the $\rho_{I,I+1}$ by
\begin{align}
  \omega_1 + \omega_2 = -2\pi i \left(1+ \frac{1}{\kappa}\sum_{I=1}^D \rho_{I,I+1} \right)\ ,
\end{align}
 where $\kappa$ is a positive integer determined by the toric data defining SE$_5$, by
 \begin{align}
  \sum_{I=1}^D (\v_I \times \v_{I+1}) = (0,0,\kappa)\ .
\end{align}
%
Then, we find
\begin{align}
  &\hspace{-25mm}\tilde{F}_{I,h^{(I)}}^{(4d)}(\omega_i, \Delta_I; u_I) \quad \nn\\
  = \prod_{\substack{a,b=1 \\ a\neq b}}^{n_I} \prod_{m=0}^{k_I-1}
  \,\,\, & \Gamma \!\left(-\frac{u_{I,ab}+\llbracket h_{I,b}-h_{I,a}-\alpha_I m \rrbracket \,\frac{\rho_{I-1,I}}{k_I} + m \frac{\rho_{I,I+1}}{k_I} - \frac{\omega_1}{2\pi i} }{\rho_{I-1,I}},-\frac{1}{\rho_{I-1,I}}, - \frac{\rho_{I,I+1}}{\rho_{I-1,I}}\right)	\nn\\
  & \times \Gamma \!\left(\frac{u_{I,ab}+\llbracket h_{I,b}-h_{I,a}-\alpha_I m \rrbracket \,\frac{\rho_{I-1,I}}{k_I} + m \frac{\rho_{I,I+1}}{k_I} - \frac{\omega_1}{2\pi i} -1}{\rho_{I,I+1}},-\frac{1}{\rho_{I,I+1}}, \frac{\rho_{I-1,I}}{\rho_{I,I+1}}\right)		\nn\\
   &  \times \Gamma \!\left(-\frac{u_{I,ab}+\llbracket h_{I,b}-h_{I,a}-\alpha_I m \rrbracket \,\frac{\rho_{I-1,I}}{k_I} + m \frac{\rho_{I,I+1}}{k_I} - \frac{\omega_2}{2\pi i} }{\rho_{I-1,I}},-\frac{1}{\rho_{I-1,I}}, - \frac{\rho_{I,I+1}}{\rho_{I-1,I}}\right)	\nn\\
   &\times  \Gamma \!\left(\frac{u_{I,ab}+\llbracket h_{I,b}-h_{I,a}-\alpha_I m \rrbracket \,\frac{\rho_{I-1,I}}{k_I} + m \frac{\rho_{I,I+1}}{k_I} - \frac{\omega_2}{2\pi i} -1}{\rho_{I,I+1}},-\frac{1}{\rho_{I,I+1}}, \frac{\rho_{I-1,I}}{\rho_{I,I+1}}\right)		\nn\\
   & \times  \Gamma \!\left(-\frac{u_{I,ab}+\llbracket h_{I,b}-h_{I,a}-\alpha_I m \rrbracket \,\frac{\rho_{I-1,I}}{k_I} + m \frac{\rho_{I,I+1}}{k_I}  }{\rho_{I-1,I}},-\frac{1}{\rho_{I-1,I}}, - \frac{\rho_{I,I+1}}{\rho_{I-1,I}}\right)^{-1}	\nn\\
   &\times  \Gamma \!\left(\frac{u_{I,ab}+\llbracket h_{I,b}-h_{I,a}-\alpha_I m \rrbracket \,\frac{\rho_{I-1,I}}{k_I} + m \frac{\rho_{I,I+1}}{k_I} -1}{\rho_{I,I+1}},-\frac{1}{\rho_{I,I+1}}, \frac{\rho_{I-1,I}}{\rho_{I,I+1}}\right)^{-1}			\nn\\
   &  \times \Gamma \!\left(-\frac{u_{I,ab}+\llbracket h_{I,b}-h_{I,a}-\alpha_I m \rrbracket \,\frac{\rho_{I-1,I}}{k_I} + m \frac{\rho_{I,I+1}}{k_I} - \frac{\omega_1+\omega_2}{2\pi i} }{\rho_{I-1,I}},-\frac{1}{\rho_{I-1,I}}, - \frac{\rho_{I,I+1}}{\rho_{I-1,I}}\right)^{-1}	\nn\\
   &\times  \Gamma \!\left(\frac{u_{I,ab}+\llbracket h_{I,b}-h_{I,a}-\alpha_I m \rrbracket \,\frac{\rho_{I-1,I}}{k_I} + m \frac{\rho_{I,I+1}}{k_I} - \frac{\omega_1+\omega_2}{2\pi i} -1}{\rho_{I,I+1}},-\frac{1}{\rho_{I,I+1}}, \frac{\rho_{I-1,I}}{\rho_{I,I+1}}\right)^{-1}	\ .
\label{eq: app 4d integrand}
\end{align}
Meanwhile,
\begin{align}
  \tilde{F}_{I,I+1}^{(2d)}(\omega_i, \Delta_I; u_I, u_{I+1}) = \prod_{a=1}^{n_I} \prod_{b=1}^{n_I+1}\,\,& \vartheta\!\left(\frac{u_a^I - u_b^{I+1}+ \frac{1}{2} (\rho_{I,I+1}-1) - \frac{\omega_1 - \omega_2}{4\pi i}}{\rho_{I,I+1}},-\frac{1}{\rho_{I,I+1}}\right)\nn\\
  &\times\vartheta\!\left(\frac{-u_a^I + u_b^{I+1}+ \frac{1}{2} (\rho_{I,I+1}-1) - \frac{\omega_1 - \omega_2}{4\pi i}}{\rho_{I,I+1}},-\frac{1}{\rho_{I,I+1}}\right)		\nn\\
  & \times  \vartheta\!\left(\frac{u_a^I - u_b^{I+1}+ \frac{1}{2} (\rho_{I,I+1}-1) - \frac{\omega_1 + \omega_2}{4\pi i}}{\rho_{I,I+1}},-\frac{1}{\rho_{I,I+1}}\right)^{-1}\nn\\
  &\times \vartheta\!\left(\frac{-u_a^I + u_b^{I+1}+ \frac{1}{2} (\rho_{I,I+1}-1) - \frac{\omega_1 + \omega_2}{4\pi i}}{\rho_{I,I+1}},-\frac{1}{\rho_{I,I+1}}\right)^{-1}	\ .
\label{eq: app 2d integrand}
\end{align}
Now, by writing the total integrand in the form $e^{-V(u_I)}$, one finds a set of saddle-point equations,
\begin{align}
  \frac{\partial V(u_I)}{\partial u_{I,a_I}} = 0 \ ,
\end{align}
 for each $I=1,\dots, D$ and $a_I=1,\dots, n_I$. A straightforward extension of the basic idea of \cite{Choi:2021rxi} gives us a way to explicitly construct a solution to these saddle-point equations, known as the `parallelogram ansatz'. We will need a slight generalisation of the method applied there\footnote{See also \cite{Kim:2024ucf}. }, which we demonstrate now. For the avoidance of doubt, all notation and definitions in the box below exist in isolation from the rest of this paper.
\vspace{1em}
 \begin{center}
\begin{mdframed}[style=exampledefault]

 We consider an integral of the form
 \begin{align}
  K = \int d^{N_1}v_1\dots d^{N_p}v_p \,e^{-V(v_1,\dots,v_k)} \ .
  \label{eq: box integral}
\end{align}
We are interested in the asymptotics at large $N_i$ for each $i=1,\dots,p$. To proceed, we seek solutions to the saddle-point equations
\begin{align}
  \frac{\partial V(v_1,\dots,v_p)}{\partial v_{i,a_i}}= 0,\qquad a_i = 1,\dots,N_i,\quad i = 1,\dots,p\ .
  \label{eq: box saddle point eqs}
\end{align}
Our interest will be in potentials of the form
\begin{align}
  V(v_1,\dots,v_p) = \sum_{i=1}^p \sum_{a_i \neq b_i} A_i(v_{i,a_i} - v_{i,b_i}) + \sum_{i< j} \sum_{a_i=1}^{N_i} \sum_{a_j=1}^{N_j} B_{ij}(v_{i,a_i} - v_{j,a_j})\ ,
  \label{eq: potential general form}
\end{align}
for some functions $A_i$ and $B_{ij}$, all of which are functions of a single variable. Let us first consider the case that the functions $B_{ij}=0$, whereby $K$ just factors into $p$ integrals. One can then directly apply the machinery of \cite{Choi:2021rxi}. In particular, we further suppose that for each $i=1,\dots, p$, the function $A_i(z)$ decomposes into periodic functions as
\begin{align}
  A_i(z) = A_i^{s_i}(z) + A_i^{t_i}(z),\qquad A_i^{s_i}(z+ s_i) = A_i^{s_i}(z), \,\, A_i^{t_i}(z+ t_i) = A_i^{t_i}(z)\ ,
\label{eq: periodicity}
\end{align}
for some complex numbers $s_i,t_i$.
Then, the saddle-point equations (\ref{eq: box saddle point eqs}) can be written as
\begin{align}
  \frac{\partial V(v_1,\dots,v_p)}{\partial v_{i,a_i}} = -\sum_{\substack{b_i=1\\b_i\neq a_i}}^{N_i} \frac{\partial}{\partial v_{i,b_i}}\left(A_i^{s_i}(v_{i,a_i} - v_{i,b_i}) + A_i^{t_i}(v_{i,a_i} - v_{i,b_i})\right)\ .
\label{eq: box saddle again}
\end{align}
Then, at large $N_i$ we adopt a continuum approximation, replacing the discrete indices $a_i,b_i=1,\dots,N_i$ with a pair of continuous variables $x_i, y_i \in (-\frac{1}{2},\frac{1}{2})$ with uniform density $\rho(x_i,y_i)=1$. The saddle-point is then given by
\begin{align}
  v_{i,a_i} \to v_i(x_i, y_i) = x_i s_i + y_i t_i\ .
\label{eq: box saddle explicit}
\end{align}
Note that it is important that $v_i$ is traceless and thus lives in $\frak{su}(N_i)$, since for the integral we are interested in studying, it is implicit that we have already integrated over the decoupled $U(1)\subset U(N)$ degrees of freedom, giving rise to a sub-leading overall factor that we have thrown away.

\para 
To see that this is indeed a saddle, note that in the continuum approximation, the saddle-point equation (\ref{eq: box saddle again}) becomes
\begin{align}
  \frac{\delta V(v_1,\dots,v_p)}{\delta v_i(\tilde{x},\tilde{y})} &= - N_i\int_{-\frac{1}{2}}^{\frac{1}{2}}dx \int_{-\frac{1}{2}}^{\frac{1}{2}}dy  \left(\frac{1}{s_i} \frac{\partial}{\partial x}A_i^{s_i}(v_i(\tilde{x},\tilde{y}) - x s_i - y t_i)\right.		\nn\\
  &\hspace{50mm}\left. + \frac{1}{t_i} \frac{\partial}{\partial y}A_i^{t_i}(v_i(\tilde{x},\tilde{y}) - x s_i -y t_i) \right)		\nn\\
  &\hspace{-20mm}= - \frac{N_i}{s_i} \int_{-\frac{1}{2}}^{\frac{1}{2}}dy \,\left(A_i^{s_i}(v_i(\tilde{x},\tilde{y})-\tfrac{1}{2}s_i - y t_i) - A_i^{s_i}(v_i(\tilde{x},\tilde{y})+\tfrac{1}{2}s_i - y t_i) \right)		\nn\\
  &\hspace{-20mm}\quad  - \frac{N_i}{t_i} \int_{-\frac{1}{2}}^{\frac{1}{2}}dx \,\left(A_i^{t_i}(v_i(\tilde{x},\tilde{y})-xs_i - \tfrac{1}{2} t_i) - A_i^{t_i}(v_i(\tilde{x},\tilde{y})-xs_i +\tfrac{1}{2} t_i) \right)		\nn\\
  &\hspace{-20mm} = 0\ ,
\label{eq: box basic manipulation}
\end{align}
by virtue of the periodicity conditions (\ref{eq: periodicity}).

\para
The generalisation we need is to a broader class of potentials, which have non-zero $B_{ij}$. In particular, we want to answer the question: under what circumstances can we turn on some $B_{ij}$ while retaining (\ref{eq: box saddle explicit}) as a saddle-point? Thankfully, we'll only need to consider the rather special scenario in which some of the periods $(s_i,t_i)$ coincide. Suppose for instance that
\begin{align}
  s_1 = s_2\ ,
\end{align}
and we turn on some $B_{12}$ such that
\begin{align}
  B_{12}(z + s_1) = B_{12}(z)\ .
  \label{eq: box B periodicity}
\end{align}
We then study the extra term in the corresponding saddle-point equation,
\begin{align}
   \frac{\partial V(v_1,\dots,v_p)}{\partial v_{1,a_1}} = - \sum_{a_2=1}^{N_2} \frac{\partial}{\partial v_{2,a_2}}B_{12}(v_{1,a_1}-v_{2,a_2})\ .
\end{align}
In the continuum limit, this can be written
\begin{align}
  \frac{\delta V(v_1,\dots,v_p)}{\delta v_1(\tilde{x},\tilde{y})} &= -\frac{N_2}{s_1} \int_{-\frac{1}{2}}^{\frac{1}{2}} dx \int_{-\frac{1}{2}}^{\frac{1}{2}} dy \, \frac{\partial}{\partial x} B_{12}(v_1(\tilde{x},\tilde{y})-xs_2 - y t_2)		\nn\\
  &\hspace{-20mm}= - \frac{N_2}{s_1} \int_{-\frac{1}{2}}^{\frac{1}{2}} dy \,\Big(B_{12}(v_1(\tilde{x},\tilde{y})-\tfrac{1}{2}s_1 - y t_2) - B_{12}(v_1(\tilde{x},\tilde{y})+\tfrac{1}{2}s_1 - y t_2)\Big)		\nn\\
   &\hspace{-20mm}= 0\ ,
\end{align}
by virtue of the periodicity condition (\ref{eq: box B periodicity}). A similar calculation shows that $ \frac{\partial V(v_1,\dots,v_p)}{\partial v_{2,a_2}} $ also still vanishes after the introduction of $B_{12}$.

\para
We can finally go to a yet more special scenario, in which
\begin{align}
  s_1 = s_2,\qquad t_1 = t_2\ .
\end{align}
Then, it is straightforward to check that we are allowed to turn on a more general $B_{12}$, taking the form
\begin{align}
  B_{12}(z) = B_{12}^{s_1}(z) + B_{12}^{t_1}(z), \quad B_{12}^{s_1}(z+s_1) = B_{12}^{s_1}(z),\,\, B_{12}^{t_1}(z+t_1) = B_{12}^{t_1}(z)\ ,
\end{align}
while not spoiling the saddle-point (\ref{eq: box saddle explicit}).

\para
In summary, we find that (\ref{eq: box saddle explicit}) is a saddle-point of the integral (\ref{eq: box integral}) if the following conditions are met:
\begin{itemize}
  \item The functions $A_i(z)$ decompose as $A_i(z) = A_i^{s_i}(z)+A_i^{t_i}(z)$ with
  \begin{align}
   A_i^{s_i}(z+ s_i) = A_i^{s_i}(z), \,\, A_i^{t_i}(z+ t_i) = A_i^{t_i}(z)\ .
\end{align}
\item If $\{s_i,t_i,s_j,t_j\}$ are all distinct, then $B_{ij}=0$. If however either $s_i = s_j = \gamma$ and $t_i\neq t_j$, or $s_i = t_j$ and $t_i \neq s_j$, or $t_i = s_j$ and $s_i \neq  t_j$, then we just require that $B_{ij}(z)$ satisfies
\begin{align}
  B_{ij}(z+\gamma) = B_{ij}(z)\ .
\end{align}
Finally, if $s_i = s_j = \gamma_1$ and $t_i = t_j = \gamma_2$, or $s_i = t_j = \gamma_1$ and $t_i = s_j = \gamma_2$, then we just require that $B_{ij}(z)$ decomposes as $B_{ij}(z) = B_{ij}^{\gamma_1}(z) + B_{ij}^{\gamma_2}(z)$ where
  \begin{align}
   B_{ij}^{\gamma_1}(z+ \gamma_1) = B_{ij}^{\gamma_1}(z), \,\, B_{ij}^{\gamma_2}(z+ \gamma_2) = B_{ij}^{\gamma_2}(z)\ .
\end{align}
\end{itemize}
If this felt a little too good to be true, it's because generically, it is. We will be interested in integrals (\ref{eq: box integral}) where the integrand $e^{-V}$ has both zeros and poles. These both correspond to branch points of the potential $V$, with an order $n$ zero (pole) giving rise to a degree\footnote{As we go around a logarithmic branch point of degree $k$, we pick up a phase $2\pi i k$.} $n$ ($-n$) logarithmic branch point. Thus, in order to define a single-valued $A_i^{s_i}(z)$, say, we need to choose some network of branch cuts; such a network better itself be periodic under $z\to z+s_i$, so as to not spoil the all-important periodicity condition on $A_i^{s_i}(z)$. Making such a choice for each component of the potential (\ref{eq: potential general form}), we arrive at a single-valued (but generically discontinuous) potential. We should think of different choices of branch cuts giving rise to different potentials $V,V'$ which have $e^{-V}= e^{-V'}$, and thus define the same matrix integral $K$.

\para
With such a broader class of potentials in mind, we must reassess the steps we have taken. The dangerous step is in going from the first to the second line of (\ref{eq: box basic manipulation}). This step still holds, precisely if as we perform the $x$ integral at fixed $y$ we do not pass through a branch cut of $A_i^{s_i}$, and similarly, as we perform the $y$ integral at fixed $x$, we do not pass through a branch cut of $A_i^{t_i}$. Otherwise, we pick up an extra term, and thus generically spoil the satisfaction of the saddle-point equation; this is bad. 

\para
It turns that there is a type of branch point that is in fact admissible. Such branch points arise when, say, $e^{-A_i^{s_i}(z)}$ has a line of single zeros at all points $z=\kappa n$ for some $\kappa\in \mathbb{C}$ and $n\in \Z$ (in addition to any other zeros and poles). One can thus pull out a factor of $\prod_{a\neq b} \left(1-e^{2\pi i \kappa(v_{i,a}-v_{i,b})}\right)$ in the matrix integral. It was shown in \cite{Choi:2021rxi}, in a computation closely related to the Molien-Weyl formula, that such a factor can be replaced in the matrix integral by $N_i! \sum_{a<b}\left(1-e^{2\pi i\kappa (v_{i,a}-v_{i,b})}\right)$ without changing the answer. It is then straightforward to show that this term is analytic on the parallelogram. In the below, it is implicit that we have already removed all admissible branch points. We are left then to worry about the scenario in which our parallelogram ansatz encounters other, inadmissible branch points.

\para
There is a simple way in which we can avoid this scenario altogether: if, as we vary the $v_i$ over their respective parallelograms (\ref{eq: box saddle explicit}), we never encounter a pole or zero of the integrand, or equivalently, a branch point of the potential. It is clear then that there exists a choice of branch cuts such that each $A_i^{s_i}, A_i^{t_i}, B_{ij}$ is single-valued, periodic and continuous on the parallelogram, and thus the saddle-point equation still holds. This was the perspective taken in \cite{Choi:2021rxi}, and for the matrix integrals of interest to us, corresponds to imposing a set of `branch point constraints' on the chemical potentials that parameterise the matrix integral, precisely so as to ensure the avoidance of all branch points. 

\para
We propose a somewhat different perspective, which allows us to explore beyond such constraints. We argue that, with a suitable configuration of branch cuts, the configuration (\ref{eq: box saddle explicit}) continues to be a saddle-point of $K$ even when one encounters branch points inside the parallelogram. In particular, we choose to connect every branch point of $A_i^{s_i}(z)$ to infinity with a branch cut that goes to infinity along the positive $s_i$ direction, and similarly, connect every branch point of $A_i^{t_i}(z)$ to infinity with a branch cut that goes to infinity along the positive $t_i$ direction. It is then simple to see that the computation (\ref{eq: box basic manipulation}) goes through unscathed, since the $x$ integral of the first term lies exactly parallel to all branch cuts, and similarly for the $y$ integral of the second term. To complete this argument, we make a similar choice of branch cuts for any non-zero interaction terms. In the case of a single shared period $s_1$, we connect all branch points of $B_{12}$ to infinity along the $s_1$ direction, while in the case of two shared periods, we connect branch points of $B_{12}^{s_1}$ to infinity in the $s_1$ direction, and branch points of $B_{12}^{t_1}$ to infinity in the $t_1$ direction. 

\para
We thus find that the parallelogram saddle continues to be a valid saddle-point, provided that we are rather careful with how we define our potential. However, we have merely passed the buck: it turns out that the aforementioned branch point constraints rear their head once more, when we take the final and important step of actually evaluating the matrix integrand at the parallelogram saddle point!

\para
In detail, the potential evaluated at the saddle-point includes a term
\begin{align}
  V|_\text{saddle}=\dots + N_i^2\int_{-1/2}^{1/2} dx_1dy_1 dx_2 dy_2 \, A^{s_i}_i(x_1 s_i + y_1 t_i - x_2 s_i - y_2 t_i) + \dots \ .
\label{eq: potential eval}
\end{align}
The integrand of the integral we are ultimately interested in is made up of a product of elliptic gamma functions and $q$-theta functions; it follows that $ A^{s_i}_i(x_1 s_i + y_1 t_i - x_2 s_i - y_2 t_i)$ is realised as a sum over terms of the form\footnote{We don't need to worry about branch points/cuts here, since we're ultimately interested in the value of $e^{-V}$ at the saddle, rather than $V$ itself.}
\begin{align}
  \log \left(1-b(y_1-y_2)e^{2\pi ic (x_1-x_2)} \right)\ ,
\label{eq: generic log contribution}
\end{align}
where $c=\pm 1$, and $b(y_1-y_2)$ is some complex number. Then, the branch point constraints are satisfied if and only if, for each such term,
\begin{align}
  |b(y_1-y_2)|<1\ ,
\end{align}
for all $(y_1-y_2)\in (-1,1)$. Conversely, whenever the branch point constraints are violated, there are a \textit{finite} number of terms of the form (\ref{eq: generic log contribution}) for each $|b(y_1-y_2)|>0$ for some values of $(y_1-y_2)\in (-1,1)$.

\para
The importance of this condition becomes clear, when we note
\begin{align}
  &\int_{-1/2}^{1/2}dx_1 \int_{-1/2}^{1/2}dx_2  \log \left(1-b(y_1-y_2)e^{2\pi ic (x_1-x_2)} \right) \nn\\
  &\qquad = \left\{ \begin{aligned}
 \,\, &0 &&\quad \text{if }|b(y_1-y_2)|<1	\\
 & 2\log|b(y_1-y_2)|	&&\quad \text{if }|b(y_1-y_2)|>1\ .
 \end{aligned}
\right.
\end{align}
One can carry out identical analysis for $A_i^{t_i}$, and similar analysis for the interaction terms $B_{ij}$. The punchline is this. Whenever the branch point constraints are satisfied, we have
\begin{align}
  V|_\text{saddle}=0\ ,
\end{align}
and hence the integral $K$ is $\mathcal{O}(1)$ at large $N_i$. In contrast, whenever some branch point constraints are violated, we generically find that the integral in (\ref{eq: potential eval}) is non-zero, and so
\begin{align}
  V|_\text{saddle} = \mathcal{O}(N_i^2)\ ,
\end{align}
where this is shorthand for terms both of the form $N_i^2$ and $N_i N_j$ for $i\neq j$. Thus, in this scenario, the matrix integral exhibits an asymptotic $\log K \sim N_i^2$ at large $N_i$.
\end{mdframed}
 \end{center}
 \vspace{1em}
 So, we need to verify that the integral (\ref{eq: integral for analysis}) is of the necessary form. We first have to partition the $(n_1+\dots + n_D)$ integration variables into $p$ sets. How we do this is dictated by the choice of holonomy. Namely, for each $I=1,\dots,D$, we reorganise the $\{u_{I,1},u_{I,2},\dots,u_{I,n_{I}}\}$ into subsets, such that the set of holonomies $h^{(I)}_{a_I}$ corresponding to each subset are equal. There are $k_I$ values the holonomy can take, and thus there are $k_I$ such subsets. Doing this for each $I$, we thus partition the integration variables into $p=\sum_{I} k_I \ge D$ sets.
 
 \para
 Suppose we then pick out one of these $p$ sets of integration variables, which without loss of generality we choose to be contained within $\{u_{I,1},\dots,u_{I,n_I}\}$ for some $I$. We need to ascribe to this set a pair of periods, which are precisely
 \begin{align}
  \rho_{I-1,I},\,\, \rho_{I,I+1}\ .
\end{align}
Note in particular that these periods do not depend on holonomy, and thus $k_I$ of the $p$ sets of integration variables have these same two matching periods. Furthermore, note that for each $I$, one of $(\rho_{I-1,I},\rho_{I,I+1})$ is the same as one of $(\rho_{I,I+1},\rho_{I+1,I+2})$. Thus, sets of integration variables arising from adjacent nodes of the toric diagram share one of their two periods. This tells us which interaction terms we are allowed in our potential.

\para
It is then not hard to show that our potential $V(u_I)$ takes the form (\ref{eq: potential general form}), with the functions $A,B$ having the necessary periodicity properties. To do so, one needs only to apply the simple fact that the elliptic gamma function obeys
\begin{align}
  \Gamma(x+1,\sigma,\tau) = \Gamma(x,\sigma,\tau)\ ,
\end{align}
and the $q$-theta function obeys
\begin{align}
  \vartheta(x+1,\tau) = \vartheta(x,\tau)\ .
\end{align}
We have therefore arrived at the desired result (\ref{eq: single holo order 1}), in the region of chemical potential space such that the relevant branch point constraints are satisfied. We leave a detailed analysis of these constraints in general to future work, and in the meantime, conclude this appendix with an analysis applicable in a particular, simple case. 

\subsubsection{Branch point constraints, in the simplest case} 

Let us say a little more about the branch cut constraints, in the simplest case of SE$_5=S^5$. In this case, if we employ `simple sum' reduction as advocated in \cite{Fujiwara:2023bdc}, the expansion (\ref{eq: GG expn intro}) takes a rather simple form. If $\I_N(\omega_i,\Delta)$ is the $U(N)$ $\N=4$ index, where we now make the rank $N$ explicit, we have
\begin{align}
  \I_N(\omega_1,\omega_2,\Delta_1,\Delta_2,\Delta_3) = \I_\infty (\omega_i,\Delta_I) \sum_{n=0}^\infty e^{-Nn \Delta_3} \tilde{\I}_n(\Delta_1,\Delta_2,\omega_1,\omega_2,-\Delta_3)\ .
\end{align}
Note that $\tilde{\I}_m$ and $\I_m$ are $U(m)$ matrix integrals with the same integrand, but defined with respect to different contour prescriptions \cite{Imamura:2022aua}\footnote{See also \cite{Ezroura:2024wmp} for a more recent discussion.}. As usual, let us assume that either choice of contour can be deformed through our saddle of interest; this allows us to study the objects appearing on either side in tandem.

\para
We are left then to understand the parallelogram ansatz approach to the asymptotics of
\begin{align}
  \I_m(-2\pi i \sigma,-2\pi i \tau,-2\pi i\nu_1,-2\pi i \nu_2, - 2\pi i \nu_3) = \int d^m u \prod_{a\neq b} \frac{\Gamma(u_{ab} + \nu_1,\sigma,\tau)\Gamma(u_{ab} + \nu_2,\sigma,\tau)}{\Gamma(u_{ab} + \nu_1+\nu_2,\sigma,\tau)\Gamma(u_{ab},\sigma,\tau)}\ .
\end{align}
We have already seen that the large $m$ growth is captured by the modular kernel (\ref{eq: 4d modular kernel explicitly}),
\begin{align}
  \log \I_m \sim -i\pi m^2 \frac{\nu_1 \nu_2 \nu_3}{\sigma \tau}\ ,
\end{align}
subject to the parallelogram saddle rendering the residual matrix integral $\mathcal{O}(1)$. It is this validity of this final statement that we would like to probe.

\para
Let us assume\footnote{For one or both of $\sigma,\tau$ lying in the lower-half-plane, one can nonetheless define $\I_m$ using the Plethystic expression (\ref{eq: Gamma def}) for $\Gamma(x,\sigma,\tau)$} $\text{Im}\,\sigma,\text{Im}\,\tau>0$, and take without loss of generality $\sigma+ \tau- \nu_1-\nu_2-\nu_3 = -1$ and $\text{Im}(\sigma/\tau)>0$. Careful analysis of the zeros and poles of (\ref{eq: app 4d integrand}), specialised to this case, reveals that the parallelogram saddle avoids branch points precisely if
\begin{align}
  \text{Im}\left(-\frac{\nu_I}{\tau}\right)>0\ ,\quad \text{Im}\left(\frac{\nu_I - 1}{\sigma}\right)>0\ ,\quad  \text{Im}\left(\frac{\tau - \nu_I}{\sigma}\right)>0\ ,\quad \text{Im}\left(\frac{(\nu_I - 1) - \sigma}{\tau}\right)>0\ ,
\label{eq: inequalities}
\end{align}
for each $I=1,2,3$, matching precisely the findings of \cite{Choi:2021rxi}. Our aim here is to simply show that this set of twelve inequalities for four independent complex variables does indeed admit a non-trivial solution space.

\para
Let us first of all treat each $(\sigma,\tau,\nu_1,\nu_2,\nu_3)$ independently, and determine the space of values that satisfy the inequalities (\ref{eq: inequalities}). We will return at the end to verify that this space has non-empty intersection with the constraint surface $\sigma+ \tau- \nu_1-\nu_2-\nu_3 = -1$.

\para 
Let us then fix $\arg\sigma = \theta_1$ and $\arg\tau= \theta_2$, where $\pi>\theta_1>\theta_2>0$. We first study the four inequalities arising from setting $I=1$ in (\ref{eq: inequalities}), which are inequalities between $\sigma,\tau$ and $\nu_1$. The first step is to determine the allowed values of $\nu_1$. A diagram is useful for this purpose; see Figure \ref{fig: butterfly}. We divide the complex plane into four `quadrants', with the lines $\{r e^{i\theta_1}:r\in \R \}$ and $\{r e^{i\theta_2}:r\in \R \}$ through the origin. This defines regions $I$ to $IV$, as labelled. Then, the inequalities (\ref{eq: inequalities}) require that $\nu_1$ lies in region $II$, while $(\nu_1-1)$ lives in region $IV$. Equivalently, we just need to place $\nu_1$ anywhere in the orange region. Let us do so. This is then sufficient to satisfy the first two inequalities, regardless of the value of $|\sigma|$ and $|\tau|$. In contrast, the final two inequalities put upper bounds on $|\sigma|$ and $|\tau|$. To see this for the third inequality, we draw a line starting at $\nu_1$ in the $\sigma$ direction. Then, $\tau$ must sit to the left of this line. Similarly, for the final inequality, we draw a line starting at $(\nu_1-1)$ in the $\tau$ direction. Then, $\sigma$ must sit to the right of this line. These two more subtle constraints are thus entirely summarised by the statement that $\sigma$ and $\tau$ are constrained to live somewhere on the solid blue and red lines, respectively. We finally repeat the same game for the second $(I=2)$ and third $(I=3)$ set of inequalities: we place $\nu_2$ and $\nu_3$ anywhere in the orange region, and then determine the corresponding upper bounds on $|\sigma|$ and $|\tau|$. In particular, the least upper bound of each of $|\sigma|$ and $|\tau|$ is evidently strictly positive. Thus, one finds a non-empty set of $(\sigma,\tau,\nu_1,\nu_2,\nu_3)$ satisfying the inequalities (\ref{eq: inequalities}) for every $\theta_1,\theta_2$, and for every $\nu_1,\nu_2,\nu_3$ in the orange region.
\begin{center}\vspace{-1em}
\begin{minipage}{0.84\textwidth}
\centering
\includegraphics[width=100mm]{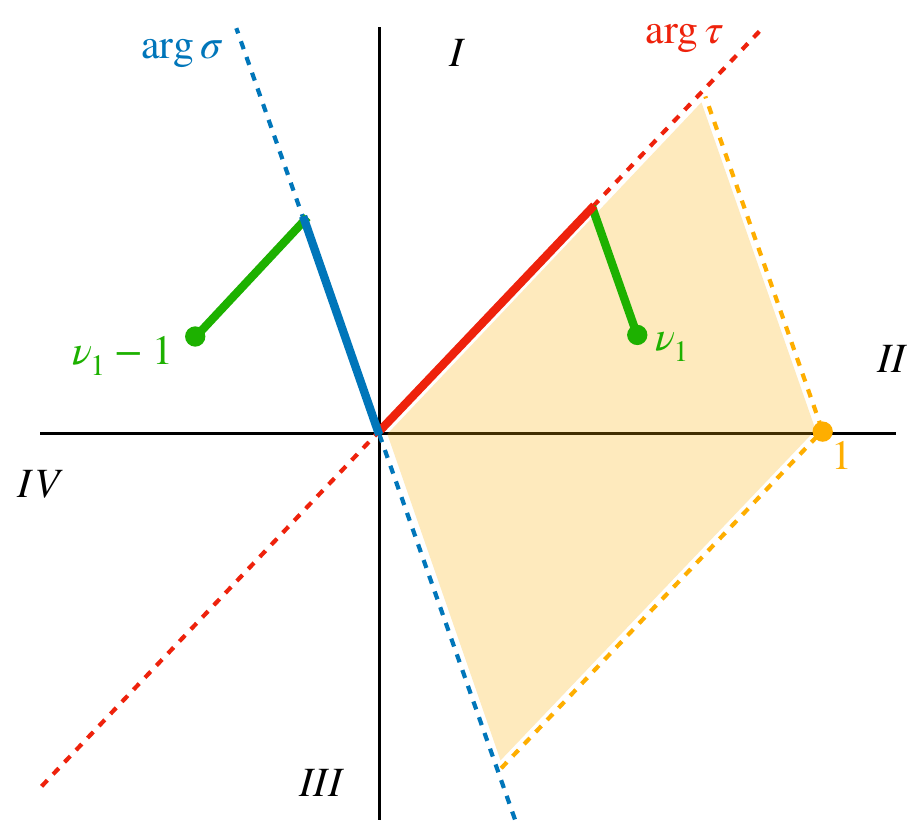}
\captionof{figure}{A diagrammatic visualisation of the solution space to the four inequalities (\ref{eq: inequalities}) with $I=1$.}\label{fig: butterfly}
\end{minipage}
\end{center}
We finally need to verify that there exist $(\sigma,\tau,\nu_1,\nu_2,\nu_3)$ satisfying the inequalities (\ref{eq: inequalities}) while also obeying the constraint $\sigma+ \tau- \nu_1-\nu_2-\nu_3 = -1$. It is convenient to parameterise the space of constraint-obeying chemical potentials by the following $8$ real variables: the arguments $\theta_1,\theta_2$ of $\sigma,\tau$, respectively, along with the real and imaginary parts of $\nu_1,\nu_2,\nu_3$. One can then compute $\sigma+\tau = \nu_1+\nu_2+\nu_3 - 1$, which in combination with $\theta_1,\theta_2$, can be used to uniquely determine $\sigma$ and $\tau$. We then verify that for each $\theta_1,\theta_2$, there exists an open subset of $(\nu_1,\nu_2,\nu_3)\in \mathbb{C}^3$ such that the inequalities are satisfied. We first exhibit a point: $\nu_1=\nu_2=\nu_3 = \frac{1}{3}(1+\epsilon e^{i(\theta_1+\theta_2)/2}) $ for real $\epsilon$, so that $\sigma+\tau = \epsilon e^{i(\theta_1+\theta_2)/2}$. It is then easy to see that there exists some $\epsilon_*>0$ such that for all $\epsilon<\epsilon_*$, the $\nu_I$ all sit in the orange region, while $|\sigma|$ and $|\tau|$ both take values beneath their least upper bounds. Finally, note that these two facts are preserved under small variations of $(\nu_1-\nu_2)$ and $(\nu_2 - \nu_3)$. Thus, we find a three-dimensional open subspace of $(\nu_1,\nu_2,\nu_3)$ space in which the inequalities are satisfied.  

\section{Computation of gauge orbit volume}
\label{app: gauge orbit}

We want to compute the volume $V(m_A)$ as defined in (\ref{eq: orbit volume def}). For this purpose it is useful to define the vectors
\begin{align}
  \w_I:= \sum_{J=1}^D \Delta_J (\v_I - \v_J)\ ,
\end{align}
which all lie in the $x-y$ plane, and satisfy $\sum_I \Delta_I \w_I = 0$, and
\begin{align}
  \w_I \times \w_J = \left(\sum_{K,L}\Delta_K \Delta_L\right) \v_I \times \v_J + \left(\sum_K \Delta_K\right) \beps \times (\v_I - \v_J)\ .
\label{eq: w cross prod}
\end{align}
The Jacobian factor is easy, 
\begin{align}
  J= \det \left(\frac{\partial (n_1,\dots,n_A)}{\partial(m_1,\dots,m_{D-2},\sigma_1,\sigma_2)}\right)  = \hatz\cdot(\w_{D-1}\times \w_D)\ ,
\end{align}
with $\hatz=(0,0,1)$, i.e. the signed area of the parallelogram with sides $\w_{D-1}$ and $\w_D$. The more fiddly part is the integral over $\sigma_1,\sigma_2$,
\begin{align}
  \iint_{P(m_A)\subset \R^2} d\sigma_1 d\sigma_2 = \text{Area}[P(m_A)]\ ,
\end{align}
which is the area (in the Euclidean metric) of a polygon in the $(\sigma_1,\sigma_2)$ plane defined by the inequalities
\begin{align}
  \u\cdot \w_A \ge - m_A,\,\, A=1,\dots,D-2,\qquad \u\cdot \w_{D-1}\ge 0,\,\, \u\cdot \w_D \ge 0,\qquad  \u = \begin{pmatrix}
  	\sigma_1\\\sigma_2\\ 0
  \end{pmatrix}\ .
\end{align}
Computing the area of this polygon, we ultimately find the familiar form
\begin{align}
  V(m_A) 	&= J\times \text{Area}[P(m_A)] \nn\\
  			&= -\frac{1}{2}\beps\cdot(\w_{D-1}\times \w_D)  \sum_{A=1}^{D-2} \left(\frac{m_A^2\,\beps \cdot (\w_{A-1}\times \w_{A+1})}{\beps \cdot (\w_{A-1}\times \w_{A})\,\,\beps \cdot (\w_{A}\times \w_{A+1})} - \frac{2m_A m_{A+1}}{\beps\cdot(\w_{A}\times \w_{A+1})}\right) \ ,
\end{align}
where the indices on the $\w_I$ are cyclic, so that $\w_{I+D}=\w_I$. The final term in this sum has only a single term. We can simplify this expression using (\ref{eq: w cross prod}), to find
\begin{align}
  V(m_A) = -\frac{1}{2}\beps\cdot(\v_{D-1}\times \v_D)  \sum_{A=1}^{D-2} \left(\frac{m_A^2\,\beps \cdot (\v_{A-1}\times \v_{A+1})}{\beps \cdot (\v_{A-1}\times \v_{A})\,\,\beps \cdot (\v_{A}\times \v_{A+1})} - \frac{2m_A m_{A+1}}{\beps\cdot(\v_{A}\times \v_{A+1})}\right) \ .
\label{eq: orbit volume result}
\end{align}
In particular, at the saddle-point we have
\begin{align}
  V(m_A^*) =  -\frac{\beps \cdot(\v_{D-1}\times \v_D)}{\omega_1 \omega_2}  \B(m_A^*) = \frac{1}{12}\left(\frac{N}{\omega_1 \omega_2}\right)^2 \beps \cdot(\v_{D-1}\times \v_D) \sum_{I,J,K}C_{I,J,K}\Delta_I \Delta_J \Delta_K\ .
  \label{eq: V at saddle}
\end{align}

\bibliography{GGandVM}
\bibliographystyle{JHEP}

\end{document}